\documentclass[aps,pre,twocolumn]{revtex4-1} 	

\usepackage{graphicx, amsmath, amssymb}
\usepackage{hyperref}
\hypersetup{ 
	colorlinks   = true
}

\newcommand \bea {\begin{eqnarray}} 
\newcommand \eea {\end{eqnarray}} 
\newcommand \be {\begin{equation}} 
\newcommand \ee {\end{equation}}

\def \>{\rangle} 
\def \<{\langle}

\begin{abstract}
	A fundamental goal of microbial ecology is to understand what determines the diversity, stability, and structure of microbial ecosystems.  The microbial context poses special conceptual challenges because of the strong mutual influences between the microbes and their chemical environment through the consumption and production of metabolites.  By analyzing a generalized consumer resource model that explicitly includes cross-feeding, stochastic colonization, and thermodynamics, we show that complex microbial communities generically exhibit a transition as a function of available energy fluxes from a ``resource-limited'' regime where community structure and stability is shaped by energetic and metabolic considerations to a diverse regime where the dominant force shaping microbial communities is the overlap between species' consumption preferences. These two regimes have distinct species abundance patterns, different functional profiles, and respond differently to environmental perturbations. Our model reproduces large-scale ecological patterns observed across multiple experimental settings such as nestedness and differential beta diversity patterns along energy gradients. We discuss the experimental implications of our results and possible connections with disorder-induced phase transitions in statistical physics.
\end{abstract}

\begin{document}
	\title{Available energy fluxes drive a transition in the diversity, stability, and functional structure of microbial communities}
	\author{Robert Marsland III}
	\email{marsland@bu.edu}
	\author{Wenping Cui} 
	\altaffiliation{Department of Physics, Boston College, Chestnut Hill, MA}
	\author{Joshua Goldford}
	\altaffiliation{Bioinformatics Program, Boston University, Boston, MA}
	\author{Alvaro Sanchez}
	\altaffiliation{Department of Ecology and Evolutionary Biology, Yale University, New Haven, CT}
	\author{Kirill Korolev}
	\author{Pankaj Mehta}
	\affiliation{Department of Physics, Boston University, Boston, MA}
	
	\date{\today}


	\maketitle

\section*{Introduction}
Microbial communities inhabit every corner of our planet, from our own nutrient-rich guts to the remote depths of the ocean floor. Different environments harbor very different levels of microbial diversity: in some samples of non-saline water at mild temperature and pH, nearly 3,000 coexisting types of bacteria can be detected, whereas at ambient temperatures warmer than 40$^\circ$ C, most cataloged samples contain fewer than 100 distinct variants \cite{EMP}.  The functional structure of these communities is also highly variable, with functional traits often reflecting the environment in which the communities are found \cite{EMP,HMP}.  A central goal of microbial community ecology is to understand how these variations in diversity, stability and functional structure \cite{Widder2016} arise from an interplay of environmental factors such as energy and resource availability \cite{Loreau1995,Embree2015} and ecological processes such as competition \cite{Gause1935,MacArthur1970,Levin1970,Chesson1990} and stochastic colonization \cite{Chase2003, Jeraldo2012, Kessler2015, Vega2017}.

This endeavor is complicated by the fact that microbes dramatically modify their abiotic environments through consumption and secretion of organic and inorganic compounds. This happens on a global scale, as in the Great Oxidation Event two billion years ago \cite{Bekker2004,Schirrmeister2015}, and also on smaller scales relevant to agriculture, industry and medicine. In this sense, every microbe is an ``ecosystem engineer'' \cite{Jones1994}. Metabolic modeling and experiments suggests that  metabolically-mediated syntrophic interactions should be a generic feature of microbial ecology \cite{Goldford2018, Harcombe2014, Zomorrodi2016} and that  complex microbial communities can self-organize even in constant environments with no spatial structure or predation \cite{Friedman2017, Goldford2018}. For these reasons, there has been significant interest in developing new models for community assembly suited to the microbial setting \cite{Taillefumier2017,Goyal2018,Good2018,Butler2018,Tikhonov2018}.

\begin{figure*}[t!]
	\includegraphics[width=17cm,trim=0 -20 0 0]{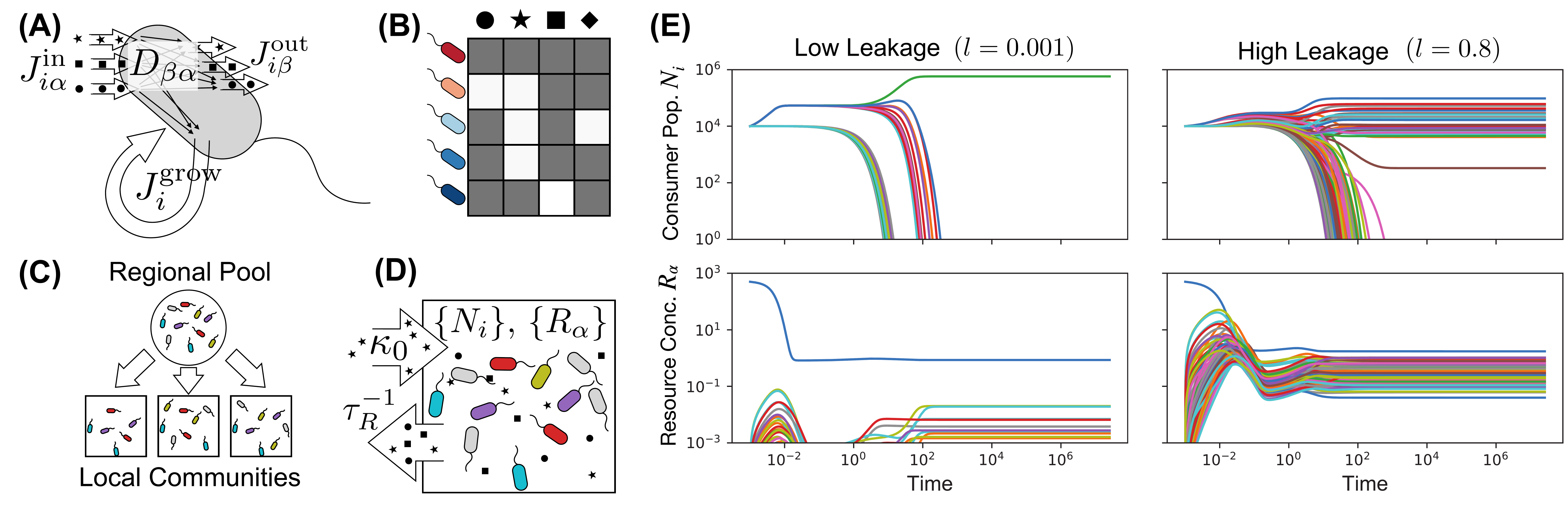}
	\caption{{\bf Microbial communities engineer complex chemical environments using a single energy source.} (A) Schematic of microbe-mediated energy fluxes in the Thermodynamic Microbial Consumer Resource Model. Each cell of species $i\, (=1,2,\dots S)$ supplies itself with energy through import of resources, generating an incoming energy flux $J_{i\alpha}^{\rm in}$ for each resource type $\alpha \,(= 1,2,\dots M)$. A fraction $l_\alpha$ of this energy leaks back into the environment in the form of metabolic byproducts, with each byproduct type carrying an outgoing energy flux $J^{\rm out}_{i\beta} = \sum_\alpha l_\alpha D_{\beta\alpha} J^{\rm in}_{i\alpha}$. The remaining energy, $J^{\rm grow}_i$, is used to replicate the cell. (B) Each species is defined by a vector of consumer preferences that encode its capacity for harvesting energy from each resource type. These vectors comprise a consumer matrix $c_{i\alpha}$. (C) A regional pool of species is randomly generated, and communities are initialized with random subsets of these species to simulate stochastic colonization. (D) Each community is supplied with a constant flux $\kappa_0$ of a single resource type ($\alpha = 0$), and all resources are continuously diluted at a fixed rate $\tau_R^{-1}$. (E) Consumer populations $N_i$ and resource concentrations $R_\alpha$ as a function of time for two realizations of this model, with low ($l = 0.001$) and high ($l = 0.8$) levels of uniform metabolic leakage (see Appendix \ref{app:procedure} for parameters).}
	\label{fig:diagram}
\end{figure*}

Here, we present a statistical physics-inspired consumer resource model for microbial community assembly that builds upon the simple model introduced in \cite{Goldford2018} and explicitly includes energetic fluxes, stochastic colonization, syntrophy, and resource competition. We focus on modeling complex communities with many species and metabolites. By necessity, any mathematical model of such a large, diverse ecosystem will contain thousands of parameters that are hard to measure. To circumvent this problem, we take a statistical physics approach where all consumer preferences and metabolic parameters are drawn from random distributions.

This approach to modeling complex systems has its root in the pioneering work of Wigner on the spectrum of heavy nuclei \cite{Wigner1955} and was  adapted by May to ecological settings \cite{May2001}. Recently, there has been a renewed interest in using these ideas to understand complex systems in both many-body physics (reviewed in \cite{dAlessio2016}) and community assembly \cite{Fisher2014, Kessler2015,  Dickens2016, Bunin2017, Biroli2017, Gibbs2017, Advani2018, Goldford2018, Barbier2018,Tikhonov2018}. The key insight underlying this approach is that generic and reproducible large-scale patterns observed across multiple settings likely reflect \emph{typical} properties, rather than fine tuned features of any particular realization or community. Consistent with this idea, it was recently shown that a generalized consumer resource model with random parameters can reproduce many of the patterns observed in experiments where natural communities were grown in synthetic minimal environments \cite{Goldford2018}.

In this paper, we ask how varying the energy flux into an ecosystem and the amount of cross-feeding affects microbial community assembly. We find that the resulting communities generically fall into two distinct regimes, characterized by qualitative differences in their community-level metabolic networks, functional structures, responses to environmental perturbations, and large-scale biodiversity patterns. We show our model predictions are consistent with data from the Tara Oceans database \cite{Sunagawa2015} and the Earth Microbiome Project \cite{EMP}, and propose feasible experimental tests using synthetic communities.

\section*{Methods}

The starting point for our analysis is a new model that adapts MacArthur's Consumer Resource Model \cite{MacArthur1970} to the microbial context by including energetics, stochastic colonization, and the exchange and consumption of metabolites. We consider the population dynamics of $S$ species of consumers (e.g., microbes) competing for $M$ types of substitutable resources. We are interested in large, diverse ecosystems where $S,M \gg 1$. A schematic summarizing our model is shown in Figure \ref{fig:diagram}.

A natural setting for considering substitutable resources is when all essential biomass components are supplied in excess, and the limiting factor for growth is the supply of usable energy. In this context, one only needs to keep track of resources from which energy can be harvested. All other nutrients are included implicitly, under the assumption that some of the energy budget is used to import whatever materials are required for growth and reproduction. Terminal waste products from which no more energy can be extracted are likewise treated implicitly, and are not included among the $M$ resource types.

In our model, the rate at which an individual of species $i$  harvests energy from resource $\alpha$ depends on the resource concentration $R_\alpha$ as well as on the consumer's vector of resource preferences $c_{i\alpha}$ through the expression:
\begin{align}
\label{eq:J}
J^{\rm in}_{i\alpha} =  w_\alpha \sigma(c_{i\alpha} R_\alpha),
\end{align}
where $\sigma(x)$ encodes the functional response and has units of mass/time, while $w_\alpha$ is the energy density of resource $\alpha$ with units of energy/mass. In the microbial context the consumer preferences $c_{i \alpha}$ can be interpreted as expression levels of transporters for each of the resources. In the main text, we focus on Type-I responses where $\sigma(x) =x$, and we set $w_\alpha = 1$ for all $\alpha$, but most of our results still hold when $\sigma(x)$ is a Monod function or the $w_\alpha$ are randomly sampled, as shown in the SI Appendix.

We model leakage and secretion by letting a fraction $l_\alpha$ of this imported energy return to the environment, so that the power available to the cell for sustaining growth is equal to
\begin{align}
J^{\rm grow}_i = \sum_\alpha (1-l_\alpha) J^{\rm in}_{i\alpha}.
\end{align}
This parameterization guarantees that the community does not spontaneously generate usable energy in violation of the Second Law of Thermodynamics. We assume that a fixed quantity $m_i$ of power per cell is required for maintenance of a steady population of species $i$, and that the per-capita growth rate is proportional to the remaining energy flux, with proportionality constant $g_i$. In typical experimental conditions, cell death is negligible, and $m_i$ is the energy harvest required for the replication rate to keep up with the dilution rate. Under these assumptions, the time-evolution of the population size $N_i$ of species $i$ can be modeled using the equation
\begin{align}
\label{dNdt}
\frac{dN_i}{dt} = g_i N_i\left[J^{\rm grow}_i - m_i\right].
\end{align}
The leaked energy flux $J^{\rm out}_i = \sum_\alpha l_\alpha J^{\rm in}_{i\alpha}$ from each cell of species $i$ is partitioned among the $M$ possible resource types via the biochemical pathways operating within the cell. We assume that all species share a similar core metabolism, encoded in a matrix $D_{\beta\alpha}$. Each element of $D_{\beta \alpha}$ specifies the fraction of leaked energy from resource $\alpha$ that is released in the form of resource $\beta$ (note that by definition, $\sum_\beta D_{\beta \alpha}=1$). Thus, in our model the resources that are excreted into the environment are intimately coupled to the resources a cell is consuming. The outgoing energy flux contained in metabolite $\beta$ is given by
\begin{align}
J^{\rm out}_{i\beta} =w_\beta \nu^{\rm out}_{i\beta} = \sum_\alpha D_{\beta\alpha}l_\alpha J^{\rm in}_{i\alpha}.
\end{align}
The dynamics of the resource concentrations depend on the incoming and outgoing mass fluxes $\nu^{\rm in}_{i\alpha} = \sigma(c_{i\alpha} R_\alpha)$ and $\nu^{\rm out}_{i\alpha}$, which are related to the energy fluxes via the energy densities $w_\alpha$. In terms of these quantities, we have
\begin{align}
\label{eq:dRdt}
\frac{dR_\alpha}{dt} &= h_\alpha + \sum_j N_j (\nu^{\rm out}_{j\alpha} - \nu^{\rm in}_{j\alpha}),
\end{align}
with $h_\alpha$ encoding the dynamics of externally supplied resources. In this manuscript, we focus on the case where the microbial communities are grown in a chemostat with a single externally supplied resource $\alpha=0$ (Figure \ref{fig:diagram}). In this case, the resource dynamics can be described by choosing  $h_\alpha = \kappa_\alpha - \tau_R^{-1} R_\alpha$, with all the $\kappa_\alpha$ set to zero except for $\kappa_0$.
These equations for $N_i$ and $R_\alpha$, along with the expressions for $J^{\rm in}_{i\alpha}$ and $J^{\rm out}_{i\alpha}$, completely specify the ecological dynamics of the model.

This model has been implemented in a freely available open-source Python package ``Community Simulator.'' The package can be downloaded from \url{https://github.com/Emergent-Behaviors-in-Biology/community-simulator}.

\section*{Results}

To assess the typical community structure and resource pool stability for ecosystems obeying Equations (\ref{eq:J})-(\ref{eq:dRdt}), we randomly generated an $M\times M$ metabolic matrix $D_{\alpha\beta}$, and a binary $S\times M$ consumer preference matrix $c_{i\alpha}$ with $S = 200$ species and $M = 100$ resources. Realizations of the randomly drawn metabolic and consumer preference matrices are shown in Figure \ref{fig:network} (top) and Figure \ref{fig:overlap} (left), respectively.

We chose $c_{i \alpha}$ so that each species had 10 preferred resource types on average, with $c_{i\alpha}=1$, while the rest of the resources were consumed at a baseline level of $c_{i\alpha} = 0.01$. The metabolic matrix $D_{\alpha\beta}$ was sampled from a Dirichlet distribution, which guarantees that all the columns sum to 1 as required by the definition of this parameter. In Appendix \ref{app:robust}, we show that the qualitative patterns we observe are unchanged if $c_{i \alpha}$ is drawn from a Gaussian or Gamma distribution, or if the $D_{\alpha\beta}$ matrix is made less sparse. The full sampling procedure is detailed in Appendix \ref{app:model}. 

We chose our units of energy flux such that the mean maintenance cost $m_i$ over all species in the regional pool is equal to 1. To break ties between species with similar consumption profiles, we added a Gaussian random offset to the $m_i$ of each species with standard deviation 0.1. In Figure \ref{fig:mi} of the Appendices, we show that these intrinsic fitness differences do not dominate the ecological dynamics, and that many species with relatively high maintenance costs are able to reach large population sizes in the steady-state communities. 

Finally, we set all the $w_\alpha$ equal to 1, and made all the leakage fractions identical, with $l_\alpha = l$ for all $\alpha$. 

To assess the amount of variability in the results, we initialized 10 different communities by seeding each one with a random subset of 100 species from the full 200-species pool. This simulates the stochastic colonization frequently observed in microbial ecosystems, where the community composition can randomly vary depending on the set of microbes this particular local environment happened to be exposed to \cite{Obadia2017}. Figure \ref{fig:diagram} shows typical dynamical trajectories in the presence of high ($l=0.8$) and extremely low leakage ($l=0.001$). 

\begin{figure}[t!]
	\includegraphics[width=8cm,trim=0 -20 0 0]{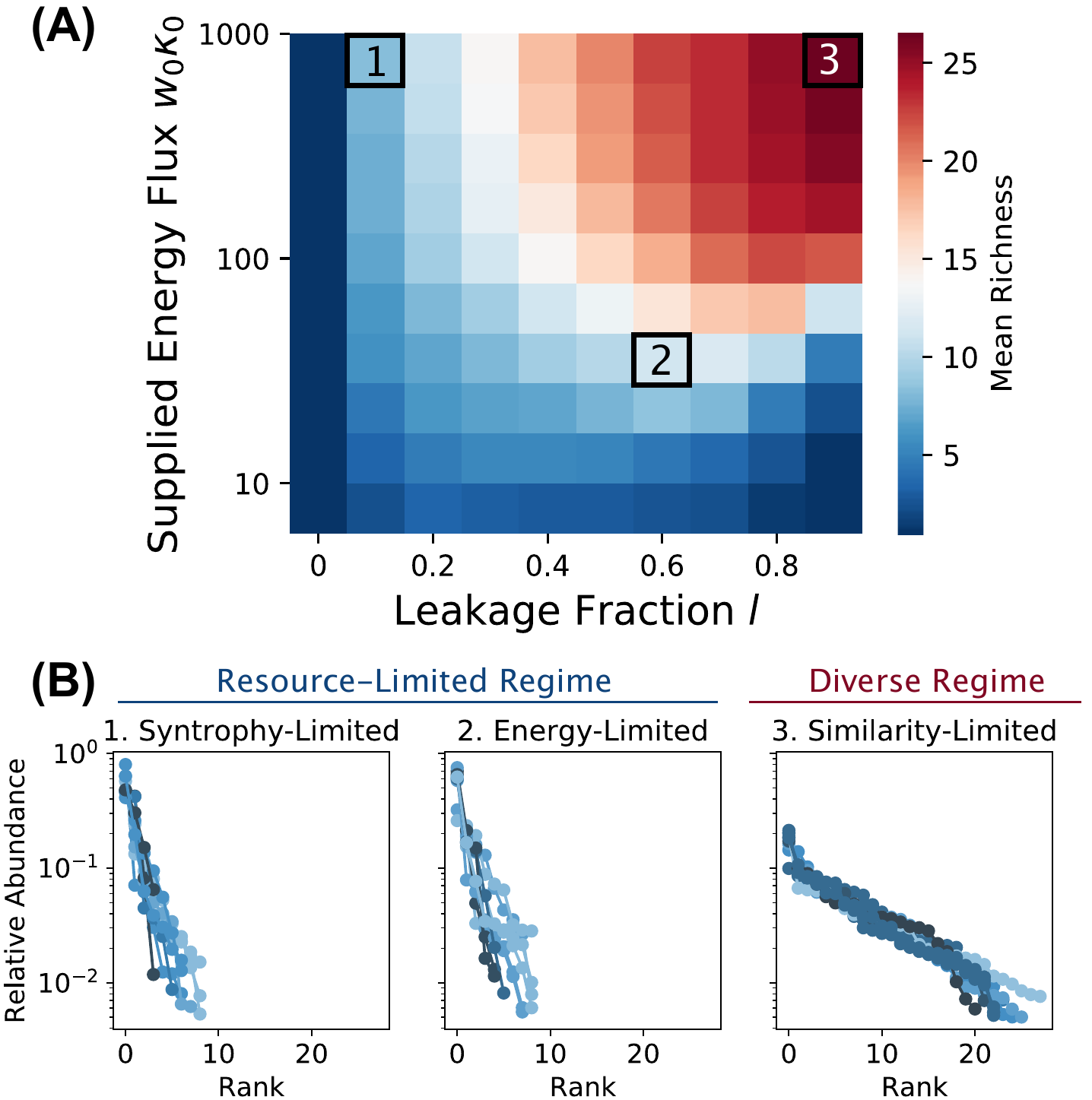}
	\caption{{\bf Steady-state richness as a function of metabolic leakage $l$ and externally supplied energy flux $w_0\kappa_0$. } We generated 200 species, initialized 10 communities of 100 species each from this pool, and ran the dynamics to steady state under different combinations of  $w_0\kappa_0$ and $l$ (see main text and Appendix \ref{app:procedure} for parameters). (A) Heat map summarizing all simulations, colored by the average number of surviving species per steady-state community (``Richness'').  Slices through the heat map are plotted in Figure \ref{fig:simpsonrichness} following the Appendices. (B) Community compositions are displayed as rank-abundance curves for three illustrative $w_0\kappa_0, l$ combinations (colored by community richness): (1) ``syntrophy-limited'' ($w_0\kappa_0 = 1000, l = 0.1$), (2) ``energy-limited'' ($w_0 \kappa_0 = 28, l = 0.6$) and (3) ``similarity-limited'' ($w_0\kappa_0 = 1000, l = 0.9$). The lines are assigned different shades for clarity. The first two examples are parts of the same resource-limited regime, manifesting similar statistical properties. The plots are truncated at a relative abundance of 0.5\%; see Figure \ref{fig:rankabundance} following the Appendices for full data.}
	\label{fig:heatmap}
\end{figure}

\subsection*{Available energy fluxes drive a transition between a ``resource-limited'' and ``diverse" regime}

Our numerical simulations display a transition between two qualitatively different community structures as we vary the externally supplied energy flux $w_0\kappa_0$ and the leakage/syntrophy $l$. In the ``thermodynamic limit'' of $M, S \to \infty$, the communities exhibit signatures of a phase transition analogous to those found in disordered systems in physics (see Discussion and Appendix \ref{app:phase}). Figure \ref{fig:heatmap} shows the effect of this transition on community diversity at our chosen finite values of $S$ and $M$. At low levels of energy flux or syntrophy, the diversity is severely limited by resource availability. In the limit of high supplied energy flux and high leakage, a maximally diverse regime is obtained, where the number of surviving species is limited only by the similarity between consumption profiles within the regional species pool, in accordance with classical niche-packing theory \cite{MacArthur1970} as we will discuss below. 

\begin{figure*}
	\includegraphics[width=17cm,trim=0 -20 0 0]{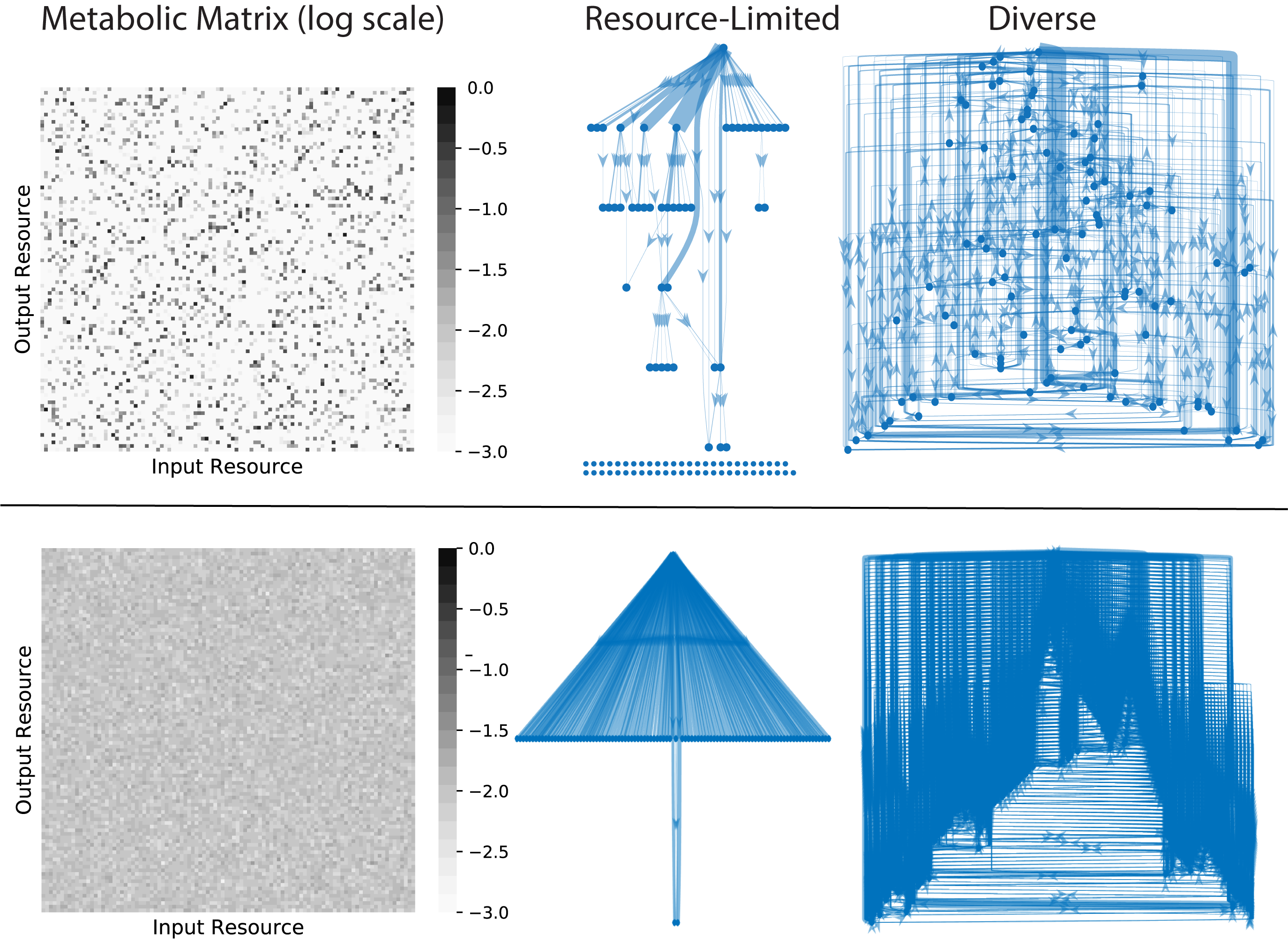}
	\caption{{\bf Energy flux networks differ in the two regimes.} Community-scale energy flux networks are plotted for a characteristic example from the diverse and resource limited regimes and two different choices of metabolic matrix $D_{\alpha\beta}$. The color of each pixel in the heat maps indicates the logarithm (base 10) of the corresponding matrix entry. In the networks, each node represents one of the $M = 100$ resource types. Edges represent steady-state energy flux from one resource type into another, mediated by consumer metabolism and leakage/secretion. The thickness of each edge is proportional to the flux magnitude, and edges with magnitudes less than 1\% of the maximum flux are not displayed. The single node at the top of each graph is the externally supplied resource, and the rows of nodes at the bottom are resources that are not connected to the external supply by any flux above the 1\% threshold. A topological analysis of the flux networks of all the simulated communities can be found in Figure \ref{fig:dag} of the Appendices.}
	\label{fig:network}
\end{figure*}

\subsection*{The resource-limited and diverse regimes produce different patterns of energy flux}

The difference between the two regimes is most apparent from the perspective of the energy flux networks. Because our model explicitly accounts for the flow of energy from one resource type into another, we can compute all the steady-state fluxes and represent them graphically, as shown in Figure \ref{fig:network} for some representative examples. Each node in this network is a resource type, and each directed edge represents the steady-state flux $J_{\beta\alpha}$ of energy conversion from resource $\alpha$ to resource $\beta$, mediated by one or more syntrophic consumers:

\begin{align}
J_{\beta\alpha} &= \sum_i N_i J^{\rm out}_{i\beta\alpha} = D_{\beta\alpha} l_\alpha \sum_i N_i w_\alpha c_{i\alpha}R_\alpha.
\end{align}

The resource-limited regime produces a unidirectional flow of energy, which is converted from the externally supplied resource type into an orderly succession of secreted resources. For the sparse metabolic matrix shown in the top row of Figure \ref{fig:network}, most resource types also have extremely small incoming flux vectors in this regime, with magnitudes less than 1\% the size of the largest flux in the network. The diverse regime displays a qualitatively different structure, where all resources have significant incoming fluxes (regardless of the choice of $D_{\alpha\beta}$), and the large number of loops in the network makes it impossible to put the resource types into any definite order. In Appendix Figure \ref{fig:dag}, we plot the fraction of samples from Figure \ref{fig:heatmap} whose (pruned) flux networks are free of cycles, and confirm that this observation is generic. The dramatic contrast between the community-level metabolism of the two regimes affects many other global features of the ecosystem, which we will explore in the following sections.

\begin{figure}
	\includegraphics[width=8cm,trim=0 -20 0 0]{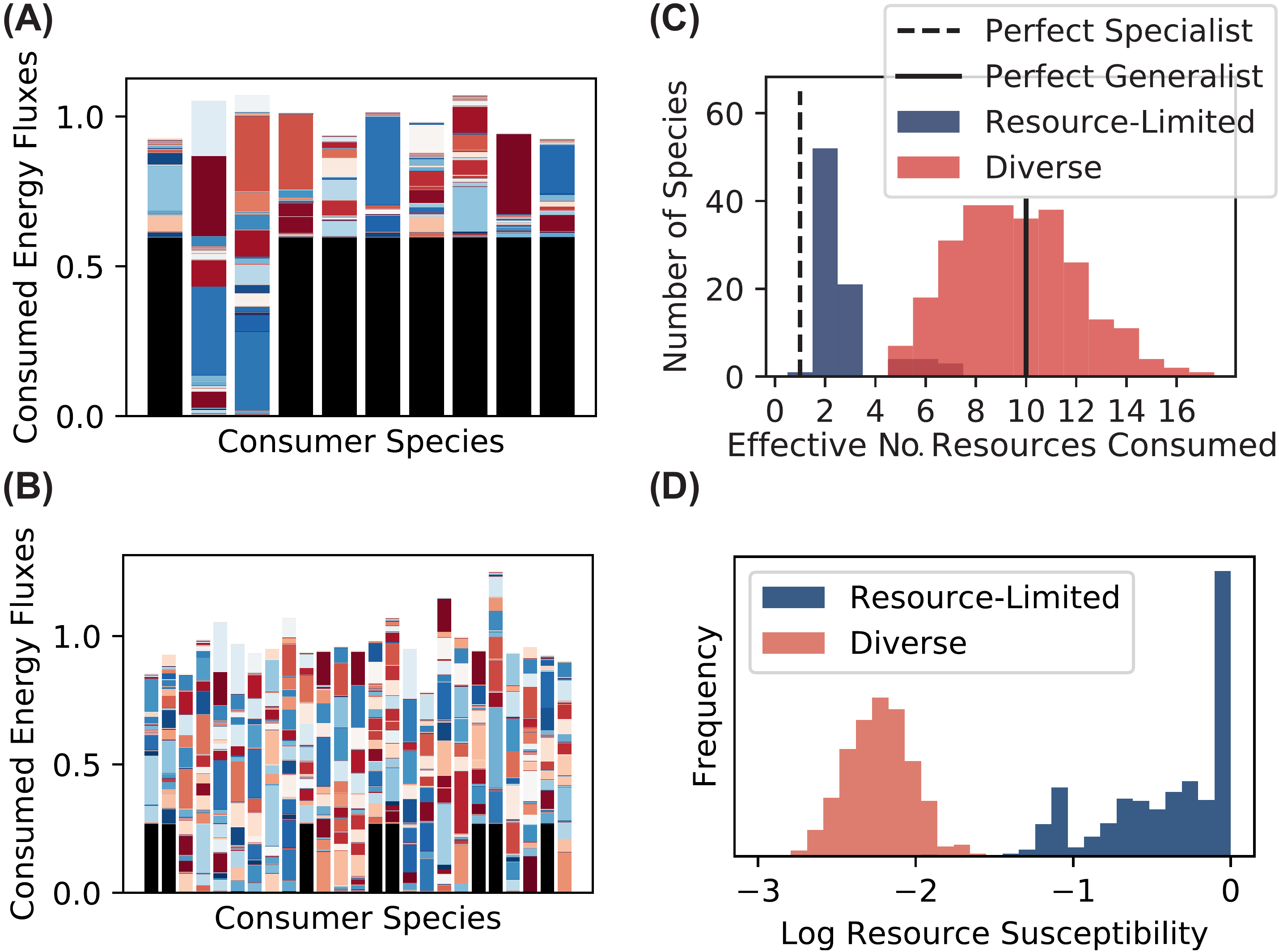}
	\caption{{\bf Structure and stability of resource dynamics depend on ecological regime}. (A) Consumed energy fluxes $(1-l)J_{i\alpha}^{\rm in}$ for each of the ten surviving species in a resource-limited community (example 2 from Figure \ref{fig:heatmap}). The black portion of the bar is the flux $(1-l)J_{i0}^{\rm in}$ due  to the externally supplied resource, and the colored bars represent the contributions of the other resources. Since these communities have reached the steady state, Equation (\ref{dNdt}) implies that the total height of each bar equals the maintenance cost $m_i$ of the corresponding consumer species. (B) Same as previous panel, but for a community from the diverse regime (example 3 from Figure \ref{fig:heatmap}). (C) Simpson diversity $M^{\rm eff}_i$ of steady-state flux vector $J^{\rm in}_{i\alpha}$ for each species from examples 2 (resource-limited) and 3 (diverse) in Figure \ref{fig:heatmap}. Vertical lines indicate the values of this metric when all the flux is concentrated on a single resource (``Specialist''), and where it is evenly spread over ten resource types (``Generalist''). (D) Logarithm of susceptibility $\log_{10} \partial \bar{R}_\alpha/\partial \kappa_\alpha$ of community-supplied resources ($\alpha \neq 0$) to addition of an externally supplied flux $\kappa_\alpha$ in these two examples.}
	\label{fig:resource}
\end{figure}

\subsection*{The two regimes have distinct functional structures}

To better understand the behavior of consumers in the two regimes, we examined the functional traits of members of typical communities in each one. In the resource-limited regime, many surviving species derive most of their energy directly from the externally supplied resource (Figure \ref{fig:resource}A).  In the diverse regime, by contrast, only a minority of the steady-state community members can consume this resource at all, and even these species receive most of their energy from a diverse array of metabolic byproducts (Figure \ref{fig:resource}B). We quantified this observation using the Simpson Diversity $M^{\rm eff}_i$ of the incoming resource flux vectors $J^{\rm in}_{i\alpha}$, which measures the effective number of resources consumed by each species, and is closely related to the inverse participation ratio in statistical physics. The Simpson Diversity is defined by
\be
M^{\rm eff}_i = \left [\sum_\alpha \left({J^{\rm in}_{i\alpha} \over J^{\rm in}_i }\right)^2 \right]^{-1},
\label{eq:simpsons}
\ee
where $J^{\rm in}_i =\sum_\alpha J^{\rm in}_{i\alpha}$ is the total incoming energy flux for each cell of this species. $M^{\rm eff}_i$ approaches 1 for species that obtain the bulk of their energy from a single resource type and approaches $M$ when all resource types are consumed equally. In the resource-limited regime, the distribution of these values is sharply peaked around 2. In the diverse regime, the peak is located around $10$, which is the average number of resources with high transporter expression in our binary sampling scheme for  $c_{i \alpha}$. This shows that most community members in the diverse regime utilize multiple energy sources, with the incoming flux spread evenly over all resource types they are capable of consuming.

\subsection*{Responses to resource perturbations differ in the two regimes}

Another important property of microbial ecosystems is how they respond to environmental perturbations. Previous theoretical studies have shown that sufficiently diverse communities can ``pin'' the resource concentrations in their local environment to fixed values, which are independent of the magnitude of externally supplied fluxes \cite{Posfai2017,Taillefumier2017,Tikhonov2017}. In these studies, resource pinning occurs only when the community saturates the diversity bound imposed by the principal of competitive exclusion, i.e. when the number of coexisting species is at least as large as the number of resource types. Such maximally diverse communities typically require fine-tuning of the resource utilization profiles or imposition of universal efficiency tradeoffs in cellular metabolism. 

In our stochastically assembled communities, the diversity is always much lower than the number of resource types, so we hypothesized that the resource concentrations should not be pinned. To test this idea, we measured the response of the steady-state concentrations $\bar{R}_\alpha$ to changes in external supply rates $\kappa_\alpha$, in terms of the ``resource susceptibilities'' $\partial \bar{R}_\alpha/\partial \kappa_\alpha$ plotted in Figure \ref{fig:resource}D \cite{Advani2018}.  Our hypothesis was valid in the resource-limited regime, where many resource susceptibilities are comparable to the susceptibility in the empty chemostat $\partial \bar{R}_\alpha/\partial \kappa_\alpha = \tau_R = 1$. But in the diverse regime, we were surprised to find that the susceptibilities are 100 times smaller than this maximum value. This suggests that resource pinning may be a generic phenomenon, observable in real ecosystems when the energy supply is sufficiently large. 
\begin{figure}
	\includegraphics[width=8cm,trim=0 -20 0 0]{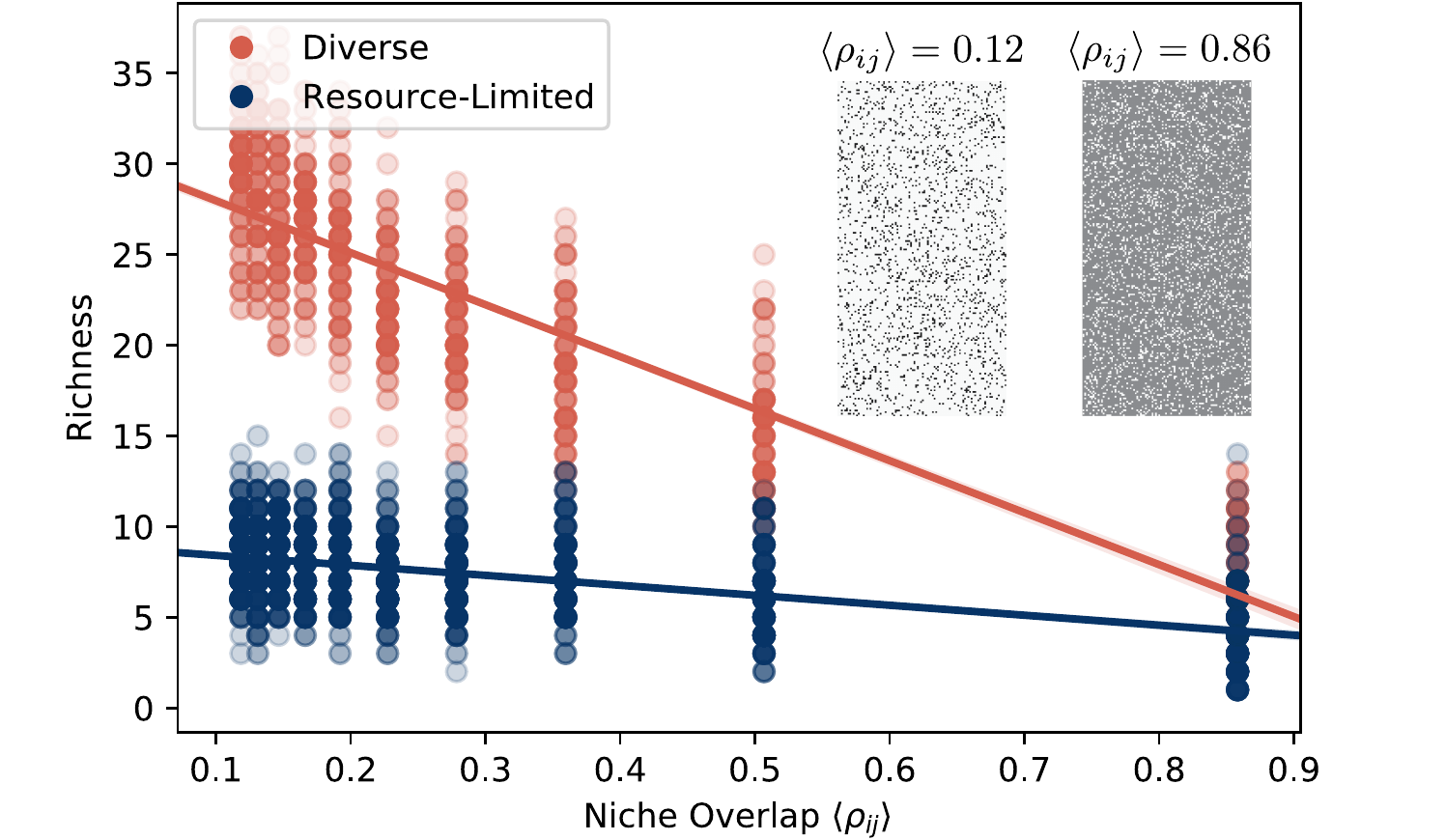}
	\caption{{\bf Richness of diverse regime depends on generalized niche-overlap}. We took the values of supplied energy flux $w_0\kappa_0$ and leakage fraction $l$ from the three examples highlighted in Figure \ref{fig:heatmap}, and varied the average niche overlap $\langle \rho_{ij}\rangle$ between members of the metacommunity. For each $w_0\kappa_0,l$ combination and each value of $\langle \rho_{ij}\rangle$, we generated 10 pools of 200 species, initialized 10 communities of 100 species each from this pool, and ran the dynamics to steady state. The steady-state richness of each community is plotted against the niche overlap. Points are colored by their regime (diverse or resource-limited), and solid lines are linear regressions. Inset: $c_{i\alpha}$ matrices that define the regional pool for two different levels of overlap, with dark squares representing high consumption coefficients.}
	\label{fig:overlap}
\end{figure}

\subsection*{Niche overlap limits richness in diverse regime}

In the diverse regime, the number of coexisting species (``richness'') is not limited by energy availability or by access to secreted metabolites, but is still much less than the maximal value of $M = 100$ set by the competitive exclusion principle \cite{Levin1970}, even though almost all $M$ resource types are present at non-negligible levels (as shown in Appendix Figure \ref{fig:resources}). We hypothesized that the diversity in this regime is limited by the degree of similarity between consumption preferences of members of the regional species pool. This can be quantified in  terms of the niche overlap \cite{MacArthur1967, Chesson1990}, whose average value in a large regional pool is given by:
\begin{align}
\langle \rho_{ij} \rangle \equiv\left\langle  \frac{\sum_\alpha c_{i\alpha}c_{j\alpha}}{\sqrt{\sum_\alpha c_{i\alpha}^2 \sum_\alpha c_{j\alpha}^2 }} \right\rangle = \frac{\langle c_{i\alpha}\rangle^2}{\sqrt{\langle c_{i\alpha}^2\rangle \langle c_{j\alpha}^2\rangle}}.
\end{align}
Figure \ref{fig:overlap} shows how the richness varies as a function of $\langle \rho_{ij}\rangle$. In the diverse regime the mean richness decreases approximately linearly with increasing overlap. The richness of the resource-limited regime, on the other hand, has only a very weak dependence on the niche overlap. These results suggest that the distribution of consumption preferences in the regional pool is the primary driver of community assembly in the diverse regime. Importantly, non-zero niche overlap limits the number of coexisting species well below the upper bound imposed by the competitive exclusion principle.

\subsection*{Nestedness and other large-scale beta-diversity patterns}

\begin{figure}[t]
	\includegraphics[width=8cm,trim=0 -20 0 0]{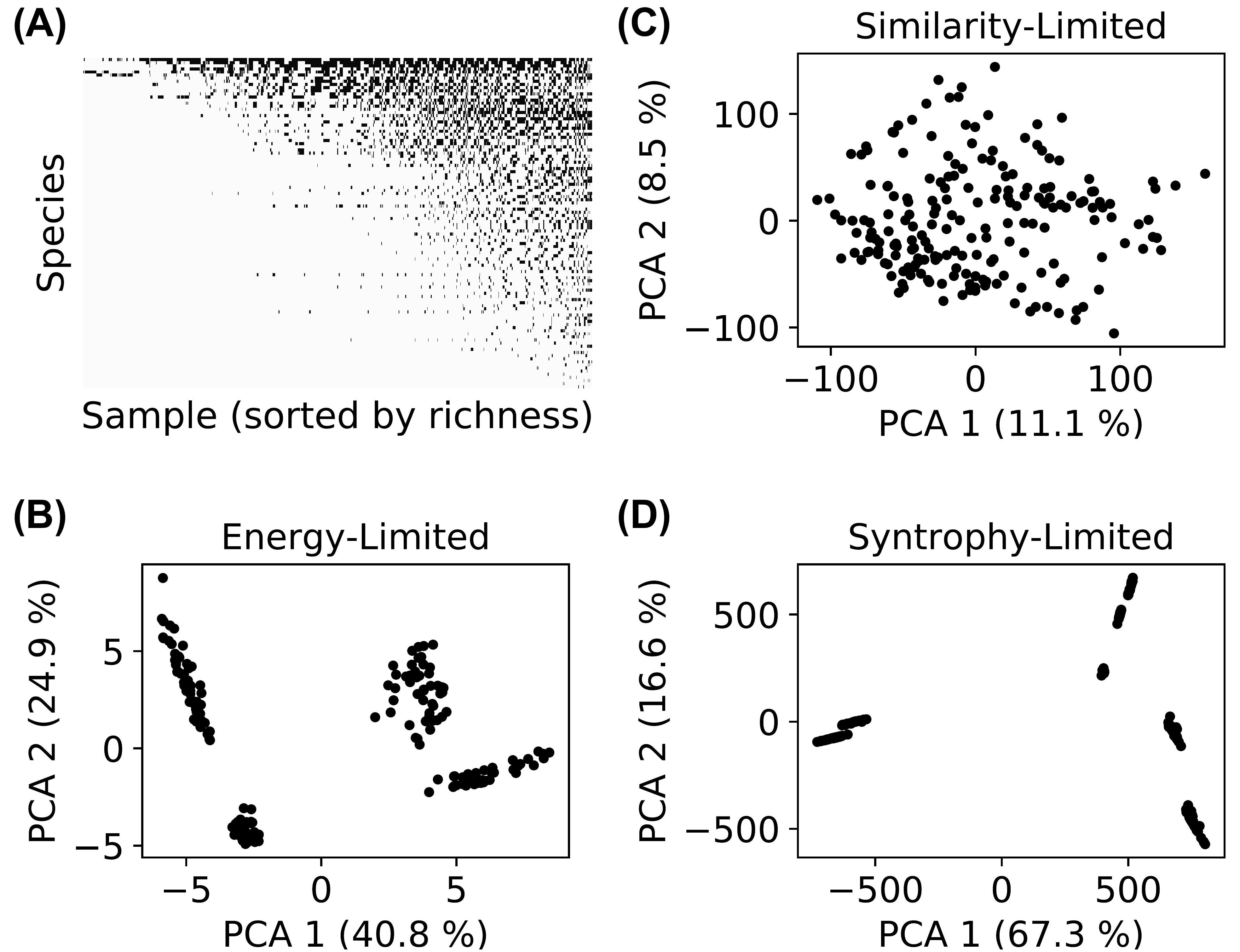}
	\caption{{\bf Resource-limited regime features community-level environmental filtering}.  (A) Presence (black) or absence (white) of all species in all 1,000 communities from the original simulations of Figure \ref{fig:heatmap}. (B, C, D) We initialized 200 new communities for each of the three examples highlighted in Figure \ref{fig:heatmap}A, by randomly choosing sets of 100 species from the regional pool. Each panel shows the projection of final community compositions $\{N_i\}$ onto the first two principal components of the set of compositions.}
	\label{fig:beta}
\end{figure}

Our aim in developing this model is to identify and understand generic patterns in community structure, that are independent of particular biological details. In large-scale surveys of natural communities, subject to many sources of noise and environmental heterogeneity, one expects that only sufficiently generic patterns will be detectable. The simplest observable to examine in such survey data is the list of species that are present or absent in each sample. We obtained these presence/absence vectors from the simulations of Figure \ref{fig:heatmap}, and found that when we sorted species by prevalence (rows in Fig. \ref{fig:beta}A)  and samples by richness (columns in Fig. \ref{fig:beta}A), the community composition generically exhibited a nested structure -- less diverse communities had a subset of the species prevalent in the more diverse communities \cite{Patterson1986,Lomolino1996}.  We quantified this result using an established nestedness metric, as described below in Appendix \ref{sec:nested}, and found that the actual nestedness exceeds the mean value for a randomized null model by more than 100 standard deviations. This suggests that nested structures may generically emerge in community assembly through the interplay of stochastic colonization, competition, and environmental filtering.

Next, we asked if we large-scale beta-diversity patterns could be used to distinguish the resource-limited and diverse regimes. We initialized 200 new communities with 100 randomly chosen members from the full regional species pool and simulated these communities to steady state in both the resource-limited and diverse regimes (see Appendix \ref{app:procedure} for details). This sub-sampling of the full regional species pools mimics the effect of stochastic colonization, where a different random subset of species seeds each community. To better understand beta-diversity signatures in the two regimes,  we performed a Principal Component Analysis (PCA) on community composition and projected the data onto the first two principal components, as shown in Figure \ref{fig:beta}B-D. In the resource-limited regime, the communities form distinct clusters that are distinguished by different highly abundant species. This suggests that harsh environments only allow a few species from the regional pool to rise to dominance, and that these dominant species induce clustering of communities.  Such ``enterotype''-like behavior is a common feature observed in many microbial settings  \cite{Arumugam2011}.  In contrast, the diverse regime exhibited neither well-defined clusters nor dominant, highly abundant species.

\subsection*{Comparison to microbial datasets}

\begin{figure}[t]
	\includegraphics[width=8cm,trim=0 -20 0 0]{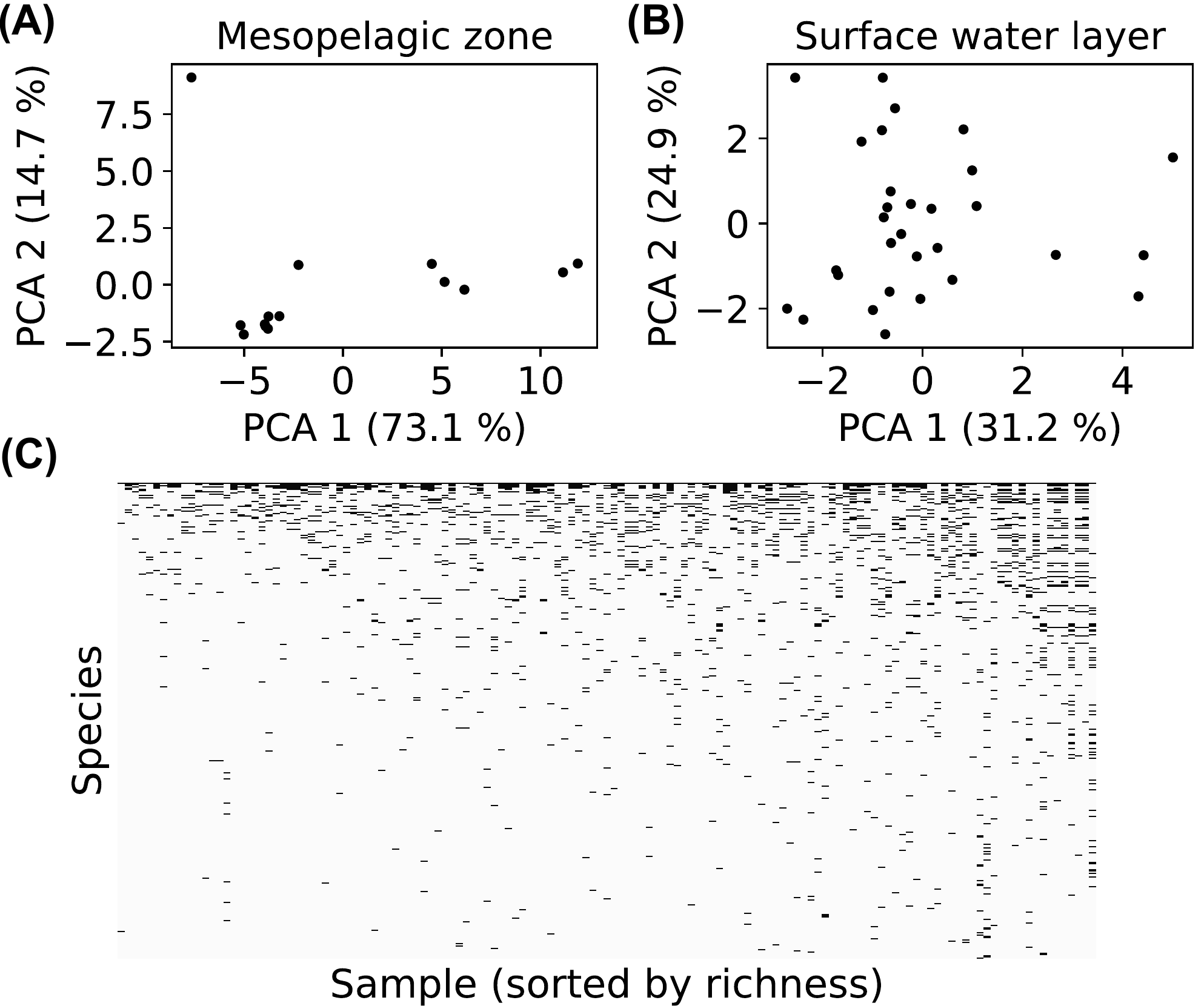}
	\caption{ {\bf Ecological regimes and nestedness in microbiome data.} (A) 16S OTU compositions of tropical mesopelagic zone samples from the Tara Oceans database, collected at a depth of 200 to 1,000 meters \cite{Sunagawa2015}. Each dot is the projection of one sample onto the first two principal components of the collection of mesopelagic zone samples. (B) Same as A, but for tropical surface water layer samples, collected at a depth of 5 meters. (C) Presence (black) or absence (white) of each OTU above 0.5\% relative abundance across all Tara Oceans samples.}
	\label{fig:data}
\end{figure}

The preceding results suggest that the resource-limited and diverse regimes can be distinguished using beta-diversity patterns. Rigorous testing of this prediction is beyond the scope of the present work. But as an illustration of the kind of data we hope to explain, we examined the natural gradient of solar energy supply in the Tara Oceans survey, which collected microbial community samples from a range of depths across the world's oceans \cite{Sunagawa2015}. Explicitly including light as an energy source would require some modification to the structure of the model equations, but we expect that the large-scale features of sufficiently diverse ecosystems should not be sensitive to changes involving just one resource.  We analyzed the 16S OTU composition of tropical ocean communities for all 30 sea-surface samples, where solar energy is plentiful, and all 13 samples from the deep-sea Mesopelagic Zone where energy fluxes are limited. We projected these composition vectors onto their first two principal components as in Figure \ref{fig:beta} above, and plot the results in Figure \ref{fig:data}. The sea surface data superficially resembles our diverse regime, with a relatively uniform distribution of possible community compositions. In contrast, the Mesopelagic Zone is similar to our resource-limited regime: the dominance of the most abundant species is much more pronounced, and the compositions appear to cluster into four discrete types. While these results are consistent with our model predictions, the number of samples at each depth is still too small to draw any definitive conclusions about clustering. 
As mentioned above, our model also gives a natural explanation for the nestedness in the Earth Microbiome Project community composition data \cite{EMP}, suggesting that it may be a natural byproduct of complex microbial communities shaped by competition, environmental heterogeneity, and stochastic colonization. To test how generic this feature is, we plotted presence/absence community compositions of all samples from the Tara Oceans dataset, sorting samples by richness and OTU's (``species'') by prevalence. Each sample contains thousands of low-abundance OTU's, which can obscure ecological patterns through their susceptibility to sequencing noise and transient immigration. We therefore imposed a 0.5\% relative abundance threshold for an OTU to count as ``present.'' The resulting pattern in Figure \ref{fig:data} is qualitatively similar to our simulations (Fig. \ref{fig:beta}D), and to the phylum-level data of the Earth Microbiome Project \cite{EMP}, with the region below the diagonal significantly less populated than the region above the diagonal. In Appendix \ref{sec:nested} and the accompanying Figure \ref{fig:nestedness-test}, we quantify the nestedness using the same metric employed in the Earth Microbiome Project analysis \cite{EMP,almeida2008consistent}, and show that the score is significantly higher than the mean scores from two standard null models. 

\section*{Discussion}
Advances in sequencing technology have opened new horizons for the study of microbial ecology, generating massive amounts of data on the composition of both natural and synthetic communities. But the complexity of these systems make it difficult to extract robust general principles suitable for guiding medical and industrial applications. Numerical simulations provide a powerful tool for addressing this problem. By rapidly iterating numerical experiments under a variety of modeling choices with random parameters, one can identify robust patterns and use the resulting insights to guide targeted experiments.

Following this strategy, we developed a thermodynamic consumer resource model that explicitly includes energetic fluxes and metabolically-mediated cross-feeding and competition. Using this model, we identified two qualitatively distinct phases in these simulations as we varied the amount of energy supplied to ecosystem and the fraction of energy leaked back into the environment: a low diversity ``resource-limited'' phase and a ``diverse'' phase. The structure of the resource-limited phase is strongly constrained by species- and community-level environmental filtering. Each community is dominated by a handful of species, making the community properties sensitive to the idiosyncratic characteristics of these species and susceptible to environmental fluctuations. In the diverse phase, communities exhibit more universal features because they substantially engineer their environments. In particular, the concentrations of resources at steady state are more narrowly distributed and insensitive to perturbations in the external supply rates. Moreover, each species draws its energy roughly equally from all resources, rather than subsisting on the externally supplied resource as in the resource-limited phase.

The emergence of environmental engineering from this community-scale model makes it a valuable tool for testing and refining existing conceptual frameworks employed by empirical biologists \cite{shou2015theory}. A major limitation of the dominant paradigms for evolution and ecology from the last century is the implicit assumption of a constant environment \cite{doebeli2017point}. The generalized Lotka-Volterra model, for example, remains a standard lens for reasoning about ecological dynamics, both quantitatively and qualitatively \cite{venturelli2018deciphering,friedman2017community,farahpour2018trade}. It assumes that the dynamics emerge from the sum of pairwise interactions among species, and that the sign and strength of these interactions are intrinsic properties of the species. This can be a good assumption in some circumstances \cite{friedman2017community,venturelli2018deciphering}, but fails to accurately describe the behavior of simple models that explicitly account for the state of the environment \cite{momeni2017lotka}. Our work provides a starting point for determining the conditions under which pairwise models will generically succeed or fail in describing the behavior of large ecosystems.

Our model complements other recent efforts at understanding microbial community ecology. Taillefumier \emph{et al.} proposed a similar model of metabolite exchange, and focused on the case where the number of resource types $M$ is equal to 3 \cite{Taillefumier2017}. In this case, repeated invasion attempts from a large regional species pool produced optimal combinations of metabolic strategies. Goyal \emph{et al.} examined the opposite limit, with $M = 5,000$, but allowed each species to consume only one type of resource \cite{Goyal2018}. This generated communities with a tree-like metabolic structure, where each species depends directly on another species to generate its unique food source. In our model, the large number of resource types ($M = 100$ in the current study) makes spontaneous strategy optimization extremely unlikely.  And our generic protocol for sampling the metabolic matrix $D_{\alpha\beta}$ allows a variety of community-level energy flux topologies to emerge, as illustrated in Figure \ref{fig:network}, which can sometimes be quite different from the tree networks of Goyal \emph{et al.}. The absence of highly specialized metabolic structure in our model makes it especially well-suited for interpreting patterns in large-scale sequence-based datasets such as the Earth Microbiome Project \cite{EMP}.

 Our model predictions can also be directly tested using experiments with natural communities in synthetic laboratory environments \cite{Datta2016, Goldford2018}. Our model predicts that beta-diversity patterns and community-level metabolic networks can be significantly altered by increasing the ecosystem's energy supply, inducing a transition from the resource-limited to the diverse regime. In the experimental set-up of \cite{Datta2016}, this can be done by directly adding chitinase enzymes to the sludge reactor to increase the degradation of chitin-based organic particles on which the ocean-derived microbial communities subsist. One could then look for shifts in the resulting diversity patterns, and observe any changes in the topology of the metabolic flux network using isotope labeling.

In this work we have largely confined ourselves to studying steady-state properties of well-mixed microbial communities. Microbial communities often exhibit complex temporal dynamics with well-defined successions \cite{Wolfe2015, Datta2016, Enke2018}. It will be interesting to explore these dynamical phenomena using our model. It is also well established that spatial structure can give rise to new ecological phenomena \cite{Korolev2010,Menon2015} and an important area of future work will be to better explore the role of space in microbial community assembly. 

Finally, we have obtained strong numerical evidence that the two regimes are separated by a phase transition, which is likely closely related to disorder-induced phase transitions in statistical physics \cite{Biroli2017}. In Appendix \ref{app:phase}, we examine the steady-state richness in the three examples of Figure \ref{fig:heatmap} under increasing values of $M$ from $M = 40$ to $M = 560$. We find that the richness is proportional to $M$ in the diverse regime, but scales sub-linearly with $M$ in both examples from the resource-limited regime. In the $M\to\infty$ limit, therefore, we expect to find a sharp line between the regimes, with the ratio of the richness to $M$ vanishing on the resource-limited side. But we do not yet know the exact location of this boundary, or the critical exponents describing the behavior of the system near the transition.

\section*{Acknowledgments}  
The funding for this work partly results from a Scialog Program sponsored jointly by Research Corporation for Science Advancement (RCSA) and the Gordon and Betty Moore Foundation. This work was also supported by NIH NIGMS grant 1R35GM119461, by a Cottrell Scholar award from RCSA to KK, and by Simons Investigator in the Mathematical Modeling of Living Systems (MMLS) awards to PM and KK. The authors are pleased to acknowledge that the computational work reported on in this paper was performed on the Shared Computing Cluster which is administered by Boston University’s Research Computing Services.  

\bibliographystyle{plain}
\bibliography{PLOS_resubmission/references.bib}

\begin{thebibliography}{10}

\bibitem{Advani2018}
Madhu Advani, Guy Bunin, and Pankaj Mehta.
\newblock {Statistical physics of community ecology: a cavity solution to
  MacArthur's consumer resource model}.
\newblock {\em Journal of Statistical Mechanics}, page 033406, 2018.

\bibitem{almeida2008consistent}
M{\'a}rio Almeida-Neto, Paulo Guimaraes, Paulo~R Guimaraes~Jr, Rafael~D Loyola,
  and Werner Ulrich.
\newblock A consistent metric for nestedness analysis in ecological systems:
  reconciling concept and measurement.
\newblock {\em Oikos}, 117(8):1227, 2008.

\bibitem{Arumugam2011}
Manimozhiyan Arumugam, Jeroen Raes, Eric Pelletier, Denis Le~Paslier, Takuji
  Yamada, Daniel~R Mende, Gabriel~R Fernandes, Julien Tap, Thomas Bruls,
  Jean-Michel Batto, et~al.
\newblock Enterotypes of the human gut microbiome.
\newblock {\em Nature}, 473:174, 2011.

\bibitem{Barbier2018}
Matthieu Barbier, Jean-Fran{\c c}ois Arnoldi, Guy Bunin, and Michel Loreau.
\newblock Generic assembly patterns in complex ecological communities.
\newblock {\em Proceedings of the National Academy of Sciences}, 2018.

\bibitem{Bekker2004}
A.~Bekker, H.~D. Holland, P.-L. Wang, D.~Rumble~III, H.~J. Stein, J.~L. Hannah,
  L.~L. Coetzee, and N.~J. Beukes.
\newblock Dating the rise of atmospheric oxygen.
\newblock {\em Nature}, 427:117, 2004.

\bibitem{Biroli2017}
Giulio Biroli, Guy Bunin, and C.~Cammarota.
\newblock Marginally stable equilibria in critical ecosystems.
\newblock {\em New Journal of Physics}, 20:083051, 2018.

\bibitem{Bunin2017}
G.~Bunin.
\newblock {Ecological communities with Lotka-Volterra dynamics}.
\newblock {\em Physical Review E}, 95:042414, 2017.

\bibitem{Butler2018}
Stacey Butler and James O'Dwyer.
\newblock Stability criteria for complex microbial communities.
\newblock {\em bioRxiv}, 2018.

\bibitem{Chase2003}
Jonathan~M Chase.
\newblock Community assembly: when should history matter?
\newblock {\em Oecologia}, 136:489, 2003.

\bibitem{Chesson1990}
Peter Chesson.
\newblock Macarthur's consumer-resource model.
\newblock {\em Theoretical Population Biology}, 37:26, 1990.

\bibitem{dAlessio2016}
Luca D'Alessio, Yariv Kafri, Anatoli Polkovnikov, and Marcos Rigol.
\newblock From quantum chaos and eigenstate thermalization to statistical
  mechanics and thermodynamics.
\newblock {\em Advances in Physics}, 65:239, 2016.

\bibitem{Datta2016}
Manoshi~S. Datta, Elzbieta Sliwerska, Jeff Gore, Martin~F. Polz, and Otto~X.
  Cordero.
\newblock Microbial interactions lead to rapid micro-scale successions on model
  marine particles.
\newblock {\em Nature Communications}, 7:11965, 2016.

\bibitem{Dickens2016}
Benjamin Dickens, Charles~K. Fisher, and Pankaj Mehta.
\newblock Analytically tractable model for community ecology with many species.
\newblock {\em Physical Review E}, 94:022423, 2016.

\bibitem{doebeli2017point}
Michael Doebeli, Yaroslav Ispolatov, and Burt Simon.
\newblock Point of view: Towards a mechanistic foundation of evolutionary
  theory.
\newblock {\em Elife}, 6:e23804, 2017.

\bibitem{Embree2015}
Mallory Embree, Joanne~K Liu, Mahmoud~M Al-Bassam, and Karsten Zengler.
\newblock Networks of energetic and metabolic interactions define dynamics in
  microbial communities.
\newblock {\em Proceedings of the National Academy of Sciences}, 112:15450,
  2015.

\bibitem{Enke2018}
Tim~N. Enke, Gabriel~E. Leventhal, Matthew Metzger, Jos\'{e}~T. Saavedra, and
  Otto~X. Cordero.
\newblock Micro-scale ecology regulates particulate organic matter turnover in
  model marine microbial communities.
\newblock {\em bioRxiv}, 2018.

\bibitem{farahpour2018trade}
Farnoush Farahpour, Mohammadkarim Saeedghalati, Verena~S Brauer, and Daniel
  Hoffmann.
\newblock Trade-off shapes diversity in eco-evolutionary dynamics.
\newblock {\em eLife}, 7:e36273, 2018.

\bibitem{Fisher2014}
Charles~K. Fisher and Pankaj Mehta.
\newblock The transition between the niche and neutral regimes in ecology.
\newblock {\em PNAS}, 111:13111, 2014.

\bibitem{Friedman2017}
Jonathan Friedman, Logan~M. Higgins, and Jeff Gore.
\newblock Community structure follows simple assembly rules in microbial
  microcosms.
\newblock {\em Nature Ecology and Evolution}, 1:0109, 2017.

\bibitem{friedman2017community}
Jonathan Friedman, Logan~M Higgins, and Jeff Gore.
\newblock Community structure follows simple assembly rules in microbial
  microcosms.
\newblock {\em Nature ecology \& evolution}, 1:0109, 2017.

\bibitem{Gause1935}
G.~F. Gause and A.~A. Witt.
\newblock Behavior of mixed populations and the problem of natural selection.
\newblock {\em The American Naturalist}, 69:596, 1935.

\bibitem{Gibbs2017}
Theo Gibbs, Jacopo Grilli, Tim Rogers, and Stefano Allesina.
\newblock The effect of population abundances on the stability of large random
  ecosystems.
\newblock {\em arXiv}, 1708.08837, 2017.

\bibitem{Goldford2018}
Joshua~E. Goldford, Nanxi Lu, Djordje Baji\'{c}, Sylvie Estrela, Mikhail
  Tikhonov, Alicia Sanchez-Gorostiaga, Daniel Segr\`{e}, Pankaj Mehta, and
  Alvaro Sanchez.
\newblock Emergent simplicity in microbial community assembly.
\newblock {\em Science}, 361:469, 2018.

\bibitem{Good2018}
Benjamin~H. Good, Stephen Martis, and Oskar Hallatschek.
\newblock Directional selection limits ecological diversification and promotes
  ecological tinkering during the competition for substitutable resources.
\newblock {\em bioRxiv}, 2018.

\bibitem{Goyal2018}
Akshit Goyal and Sergei Maslov.
\newblock {Diversity, stability, and reproducibility in stochastically
  assembled microbial ecosystems}.
\newblock {\em Physical Review Letters}, 120:158102, 2018.

\bibitem{Harcombe2014}
William~R. Harcombe, William~J. Riehl, Ilija Dukovski, Brian~R. Granger, Alex
  Betts, Alex~H. Lang, Gracia Bonilla, Amrita Kar, Nicholas Leiby, Pankaj
  Mehta, Christopher~J. Marx, and Daniel Segr\`{e}.
\newblock Metabolic resource allocation in individual microbes determines
  ecosystem interactions and spatial dynamics.
\newblock {\em Cell Reports}, 7:1104, 2014.

\bibitem{HMP}
Curtis Huttenhower, Dirk Gevers, Rob Knight, Sahar Abubucker, Jonathan~H
  Badger, Asif~T Chinwalla, Heather~H Creasy, Ashlee~M Earl, Michael~G
  FitzGerald, Robert~S Fulton, et~al.
\newblock Structure, function and diversity of the healthy human microbiome.
\newblock {\em Nature}, 486:207, 2012.

\bibitem{Jeraldo2012}
Patricio Jeraldo, Maksim Sipos, Nicholas Chia, Jennifer~M Brulc, A~Singh
  Dhillon, Michael~E Konkel, Charles~L Larson, Karen~E Nelson, Ani Qu,
  Lawrence~B Schook, F~Yang, Bryan~A White, and Nigel Goldenfeld.
\newblock Quantification of the relative roles of niche and neutral processes
  in structuring gastrointestinal microbiomes.
\newblock {\em Proceedings of the National Academy of Sciences}, 109:9692,
  2012.

\bibitem{Jones1994}
Clive~G. Jones, John~H. Lawton, and Shachak Moshe.
\newblock Organisms as ecosystem engineers.
\newblock {\em Oikos}, 69:373, 1994.

\bibitem{Kessler2015}
David~A. Kessler and Nadav~M. Shnerb.
\newblock Generalized model of island biodiversity.
\newblock {\em Physical Review E}, 91:042705, 2015.

\bibitem{Korolev2010}
Kirill~S Korolev, Mikkel Avlund, Oskar Hallatschek, and David~R Nelson.
\newblock Genetic demixing and evolution in linear stepping stone models.
\newblock {\em Reviews of Modern Physics}, 82:1691, 2010.

\bibitem{Levin1970}
Simon~A Levin.
\newblock Community equilibria and stability, and an extension of the
  competitive exclusion principle.
\newblock {\em The American Naturalist}, 104:413, 1970.

\bibitem{Lomolino1996}
Mark~V. Lomolino.
\newblock Investigating causality of nestedness of insular communities:
  Selective immigrations or extinctions?
\newblock {\em Journal of Biogeography}, 23:699, 1996.

\bibitem{Loreau1995}
Michel Loreau.
\newblock Consumers as maximizers of matter and energy flow in ecosystems.
\newblock {\em The American Naturalist}, 145:22, 1995.

\bibitem{MacArthur1970}
Robert MacArthur.
\newblock {Species Packing and Competitive Equilibrium for Many Species}.
\newblock {\em Theoretical Population Biology}, 1:1, 1970.

\bibitem{MacArthur1967}
Robert MacArthur and Richard Levins.
\newblock {The Limiting Similarity, Convergence, and Divergence of Coexisting
  Species}.
\newblock {\em The American Naturalist}, 101:377, 1967.

\bibitem{May2001}
Robert~M. May.
\newblock {\em Stability and complexity in model ecosystems}.
\newblock Princeton University Press, Princeton, N.J., 2001.

\bibitem{Menon2015}
Rajita Menon and Kirill~S Korolev.
\newblock Public good diffusion limits microbial mutualism.
\newblock {\em Physical Review Letters}, 114:168102, 2015.

\bibitem{momeni2017lotka}
Babak Momeni, Li~Xie, and Wenying Shou.
\newblock Lotka-volterra pairwise modeling fails to capture diverse pairwise
  microbial interactions.
\newblock {\em Elife}, 6:e25051, 2017.

\bibitem{Nishimori}
Hidetoshi Nishimori.
\newblock {\em Statistical Physics of Spin Glasses and Information Processing}.
\newblock Oxford University Press, New York, NY, 2001.

\bibitem{Obadia2017}
Benjamin Obadia, Z.T. G\"{u}vener, Vivian Zhang, Javier~A. Ceja-Navarro,
  Eoin~L. Brodie, William~W. Ja, and William~B. Ludington.
\newblock Probabilistic invasion underlies natural gut microbiome stability.
\newblock {\em Current Biology}, 27:1999, 2017.

\bibitem{Patterson1986}
Bruce~D Patterson and Wirt Atmar.
\newblock Nested subsets and the structure of insular mammalian faunas and
  archipelagos.
\newblock {\em Biological Journal of the Linnean Society}, 28:65, 1986.

\bibitem{scikit-learn}
F.~Pedregosa, G.~Varoquaux, A.~Gramfort, V.~Michel, B.~Thirion, O.~Grisel,
  M.~Blondel, P.~Prettenhofer, R.~Weiss, V.~Dubourg, J.~Vanderplas, A.~Passos,
  D.~Cournapeau, M.~Brucher, M.~Perrot, and E.~Duchesnay.
\newblock Scikit-learn: Machine learning in {P}ython.
\newblock {\em Journal of Machine Learning Research}, 12:2825--2830, 2011.

\bibitem{Posfai2017}
Anna Posfai, Thibaud Taillefumier, and Ned~S. Wingreen.
\newblock Metabolic trade-offs promote diversity in a model ecosystem.
\newblock {\em Physical Review Letters}, 118:028103, 2017.

\bibitem{Schirrmeister2015}
Bettina~E. Schirrmeister, Muriel Gugger, and Philip C.~J. Donoghue.
\newblock Cyanobacteria and the great oxidation event: Evidence from genes and
  fossils.
\newblock {\em Palaeontology}, 58:769, 2015.

\bibitem{shou2015theory}
Wenying Shou, Carl~T Bergstrom, Arup~K Chakraborty, and Frances~K Skinner.
\newblock Theory, models and biology.
\newblock {\em Elife}, 4:e07158, 2015.

\bibitem{Sunagawa2015}
Shinichi Sunagawa, Luis~Pedro Coelho, Samuel Chaffron, Jens~Roat Kultima,
  Karine Labadie, Guillem Salazar, Bardya Djahanschiri, Georg Zeller, Daniel~R.
  Mende, Adriana Alberti, Francisco~M. Cornejo-Castillo, Paul~I. Costea,
  Corinne Cruaud, Francesco d{\textquoteright}Ovidio, Stefan Engelen, Isabel
  Ferrera, Josep~M. Gasol, Lionel Guidi, Falk Hildebrand, Florian Kokoszka,
  Cyrille Lepoivre, Gipsi Lima-Mendez, Julie Poulain, Bonnie~T. Poulos, Marta
  Royo-Llonch, Hugo Sarmento, Sara Vieira-Silva, C{\'e}line Dimier, Marc
  Picheral, Sarah Searson, Stefanie Kandels-Lewis, Chris Bowler, Colomban
  de~Vargas, Gabriel Gorsky, Nigel Grimsley, Pascal Hingamp, Daniele Iudicone,
  Olivier Jaillon, Fabrice Not, Hiroyuki Ogata, Stephane Pesant, Sabrina
  Speich, Lars Stemmann, Matthew~B. Sullivan, Jean Weissenbach, Patrick
  Wincker, Eric Karsenti, Jeroen Raes, Silvia~G. Acinas, and Peer Bork.
\newblock {Structure and function of the global ocean microbiome}.
\newblock {\em Science}, 348:1261359, 2015.

\bibitem{Taillefumier2017}
Thibaud Taillefumier, Anna Posfai, Yigal Meir, and Ned~S. Wingreen.
\newblock {Microbial consortia at steady supply}.
\newblock {\em eLife}, 6:e22644, 2017.

\bibitem{EMP}
Luke~R. Thompson, Jon~G. Sanders, Daniel McDonald, Amnon Amir, Joshua Ladau,
  Kenneth~J. Locey, Robert~J. Prill, Anupriya Tripathi, Sean~M. Gibbons, Gail
  Ackermann, Jose~A. Navas-Molina, Stefan Janssen, Evguenia Kopylova, Yoshiki
  V\'{a}zquez-Baeza, Antonio Gonz\'{a}lez, James~T. Morton, Siavash Mirarab,
  Zhenjiang Zech~Xu, Lingjing Jiang, Mohamed~F. Haroon, Jad Kanbar, Qiyun Zhu,
  Se~Jin~Song, Tomasz Kosciolek, Nicholas~A. Bokulich, Joshua Lefler, Colin~J.
  Brislawn, Gregory Humphrey, Sarah~M. Owens, Jarrad Hampton-Marcell, Donna
  Berg-Lyons, Valerie McKenzie, Noah Fierer, Jed~A. Fuhrman, Aaron Clauset,
  Rick~L. Stevens, Ashley Shade, Katherine~S. Pollard, Kelly~D. Goodwin,
  Janet~K. Jansson, Jack~A. Gilbert, Rob Knight, and Earth Microbiome~Project
  Consortium.
\newblock {A communal catalogue reveals Earth's multiscale microbial
  diversity}.
\newblock {\em Nature}, 551:457, 2017.

\bibitem{Tikhonov2017}
Mikhail Tikhonov and Remi Monasson.
\newblock Collective phase in resource competition in a highly diverse
  ecosystem.
\newblock {\em Physical Review Letters}, 118:048103, 2017.

\bibitem{Tikhonov2018}
Mikhail Tikhonov and Remi Monasson.
\newblock {Innovation Rather than Improvement: A Solvable High-Dimensional
  Model Highlights the Limitations of Scalar Fitness}.
\newblock {\em Journal of Statistical Physics}, 2018.

\bibitem{Tuomisto2010}
Hanna Tuomisto.
\newblock {A consistent terminology for quantifying species diversity? Yes, it
  does exist}.
\newblock {\em Oecologia}, 164:853, 2010.

\bibitem{Vega2017}
Nicole~M. Vega and Jeff Gore.
\newblock Stochastic assembly produces heterogeneous communities in the
  \emph{{C}aenorhabditis elegans} intestine.
\newblock {\em PLoS Biol.}, 15:e2000633, 2017.

\bibitem{venturelli2018deciphering}
Ophelia~S Venturelli, Alex~V Carr, Garth Fisher, Ryan~H Hsu, Rebecca Lau,
  Benjamin~P Bowen, Susan Hromada, Trent Northen, and Adam~P Arkin.
\newblock Deciphering microbial interactions in synthetic human gut microbiome
  communities.
\newblock {\em Molecular systems biology}, 14:e8157, 2018.

\bibitem{Widder2016}
Stefanie Widder, Rosalind~J Allen, Thomas Pfeiffer, Thomas~P Curtis, Carsten
  Wiuf, William~T Sloan, Otto~X Cordero, Sam~P Brown, Babak Momeni, Wenying
  Shou, Helen Kettle, Harry~J Flint, Andreas~F Haas, B\'{e}atrice Laroche,
  Jan-Ulrich Kreft, Tobias Gro\ss{}kopf, Jef Huisman, Andrew Free, Cristian
  Picioreanu, Christopher Quince, Isaac Klapper, Simon Labarthe, Barth~F.
  Smets, Harris Wang, and Orkun~S Soyer.
\newblock Challenges in microbial ecology: building predictive understanding of
  community function and dynamics.
\newblock {\em The ISME journal}, 10:2557, 2016.

\bibitem{Wigner1955}
Eugene~P. Wigner.
\newblock Characteristic vectors of bordered matrices with infinite dimensions.
\newblock {\em Annals of Mathematics (ser. 2)}, 62:548, 1955.

\bibitem{Wolfe2015}
Benjamin~E. Wolfe and Rachel~J. Dutton.
\newblock Fermented foods as experimentally tractable microbial ecosystems.
\newblock {\em Cell}, 161:49, 2015.

\bibitem{Zomorrodi2016}
Ali~R. Zomorrodi and Daniel Segr\`{e}.
\newblock Synthetic ecology of microbes: Mathematical models and applications.
\newblock {\em J. Mol. Biol.}, 428:837, 2016.

\end{thebibliography}

\appendix
\part*{SI Appendices}

In Appendix \ref{app:model}, we provide a full explanation of the Microbial Consumer Resource Model. We describe its numerical implementation in Appendix \ref{app:procedure}, and present additional data illustrating the robustness of the qualitative results in Appendix \ref{app:robust}. Finally, Appendix \ref{app:phase} contains preliminary evidence for the existence of a bona fide phase transition in the $M\to\infty$ limit. All data and code for generating figures can be found at \url{https://github.com/Emergent-Behaviors-in-Biology/crossfeeding-transition}.

\section{Model details}
\label{app:model}

\subsection{Generalities}

We begin by defining an \emph{energy flux} into a cell $J^{\mathrm{in}}$, an energy flux that is used for growth $J^{\mathrm{growth}}$, and an outgoing energy flux due to byproduct secretion $J^{\mathrm{out}}$. Energy conservation requires
\be
\label{eq:energy}
J^{\mathrm{in}}=J^{\mathrm{growth}}+J^{\mathrm{out}}
\ee
for any reasonable metabolic model. 
Now consider a model with $M$ resources $R_\beta$ with $\beta=1 \ldots M$ each with ``energy'' or quality $w_\beta$. It will be useful to divide the input and output energy fluxes that are consumed/secreted in metabolite $\beta$ by $J_{\beta}^{\mathrm{in}}$ and $J_{\beta}^{\mathrm{out}}$ respectively.
We define the fraction $f_\beta^\mathrm{out}$ of the  \emph{output energy} secreted as resource $\beta$ by
\be
J_{\beta}^{\mathrm{out}} \equiv f_\beta^\mathrm{out} J^{\mathrm{out}}.
\ee
We can define corresponding mass fluxes by 
\be
\nu_{\beta}^{\mathrm{out}}\equiv J_{\beta}^{\mathrm{out}}/w_\beta
\ee
and
\be
\nu_{\beta}^{\mathrm{in}}\equiv J_{\beta}^{\mathrm{in}}/w_\beta
\ee
In general, all these fluxes depend on the species under consideration and will carry an extra roman index $i$ indicating the species.

We assume that a fixed quantity $m_i$ of power per cell is required for maintenance of species $i$, and that the per-capita growth rate is proportional to the remaining energy flux $(J^{\rm growth}-m_i)$, with proportionality constant $g_i$. Under these assumptions, the time-evolution of the population size $N_i$ of species $i$ can be modeled using the equation
\be
{dN_i \over dt}= g_i N_i (J_i^{\mathrm{growth}} -m_i).
\ee
We can model the resource dynamics by functions of the form
\be
{dR_\alpha \over dt} =h_\alpha (R_\alpha) - \sum_j N_j \nu_{j \alpha}^{\mathrm{in}} + \sum_j N_j \nu_{j \alpha}^{\mathrm{out}},
\ee
where the function $h_\alpha$ describes the resource dynamics in the absence of consumers. We can consider two kinds of dynamics: externally supplied and self-renewing. For externally supplied resources, we take a linearized form of the dynamics:
\be
h_\alpha^{\mathrm{external}} (R_\alpha) = \kappa_\alpha- \tau_\alpha^{-1} R_\alpha
\ee
while for self-renewing we take a logistic form for the dynamics
\be
h_\alpha^{\mathrm{self-renewing}} (R_\alpha) = r_\alpha R_\alpha (K_\alpha-R_\alpha).
\ee
In the present study, we only consider externally supplied resources.

These equations specify the general dynamics of all the models we consider. Metabolism is encoded in the relationship between input, output, and growth fluxes.

\subsection{Input fluxes and output partitioning}
We will now specify the form of the input fluxes $\nu_{\beta}^{\mathrm{in}}$ and the output partitioning $f_\beta^{\rm out}$. This involves specifying how an input resource is turned into an metabolic byproducts. To try to capture metabolic structure, we will divide the $M$ resources into $T$ classes (e.g. sugars, amino acids, etc.), each with $M_A$ resources where $A=1, \ldots T$ and $\sum_A M_A=M$. We will be interested in capturing coarse metabolic structure (i.e. metabolizing sugars outputs carboxylic acids, etc). We will limit ourselves to considering strictly substitutable resources.

In all consumer resource models, we assume that
\be
\nu_{i\beta}^{\mathrm{in}} = \sigma(c_{i \beta} R_\beta)
\ee
where $\sigma$ encodes the response function of consumer $i$ for resource $\alpha$. In the microbial context the consumer preferences $c_{i \alpha}$ can be interpreted as expression levels of transporters for each of the resources. We consider three kinds of response functions: Type-I, linear response functions where 
\be
\sigma_I(x)=x,
\ee
a Type-II saturating Monod function,
\be
\sigma_{II}(x) = {x \over 1+{x \over K}}
\ee
and a Type-III Hill or sigmoid-like function
\be
\sigma_{III}(x) = {x^n \over 1 +\left({x \over K}\right)^n},
\ee
where $n>1$.

In all the simulations of this paper, we assume that resources independently contribute to the growth rate.  We define a leakage fraction $0 \le l_\alpha \le 1$ for resource $\alpha$ such that
\be
J_{\alpha}^{\mathrm{out}}= l_\alpha J_{\alpha}^{\mathrm{in}}.
\ee
A direct consequence of energy conservation (Equation (\ref{eq:energy})) is that
\be
J_i^{\mathrm{growth}}= \sum_\alpha (1-l_\alpha) J_{i \alpha}^{\mathrm{in}} = \sum_\alpha (1-l_\alpha) w_\alpha \sigma(c_{i \alpha} R_\alpha)
\ee
All that is left is to determine how to compute the probability of producing a byproduct $\beta$ when consuming $\alpha$. Let us denote by $D_{\beta \alpha}$ the fraction of the output energy that is contained in metabolite $\beta$ when a cell consumes $\alpha$. Note that by definition $\sum_{\beta} D_{\beta\alpha} = 1$. These $D_{\beta \alpha}$ and $l_\alpha$ uniquely specify the metabolic model for independent resources and we can write all fluxes in terms of these quantities.  

The total energy output in metabolite $\beta$ is thus 
\be
J_{i\beta}^{\mathrm{out}} = \sum_{\alpha} D_{\beta \alpha} l_\alpha J_{i\alpha}^{\mathrm{in}}= \sum_{\alpha} D_{\beta \alpha} l_\alpha w_\alpha \sigma(c_{i \alpha} R_\alpha).
\ee
This also yields
\be
\nu_{i\beta}^{\mathrm{out}} =  \sum_{\alpha} D_{\beta \alpha} l_\alpha  {w_\alpha \over w_\beta} \sigma(c_{i \alpha} R_\alpha)
\ee
We are now in position to write down the full dynamics in terms of these quantities:
\bea
{dN_i \over dt}&=& g_i N_i \left[\sum_\alpha  (1-l_\alpha) w_\alpha \sigma(c_{i \alpha} R_\alpha) -m_i\right] \nonumber\\
{dR_\alpha \over dt} &=& h_\alpha (R_\alpha) - \sum_j N_j \sigma(c_{j \alpha} R_\alpha) \nonumber\\
&&\,\,\,\,\,\,\,\,\,\,\,\,+  \sum_{j \beta} N_j   \sigma(c_{j \beta} R_\beta) \left[ D_{\alpha \beta}{w_\beta \over w_\alpha} l_\beta \right]
\label{eq:dynamics}
\eea
Notice that when $\sigma$ is Type-I (linear) and $l_\alpha =0$ for all $\alpha$ (no leakage or byproducts), this reduces to MacArthur's original model  \cite{MacArthur1970}.

\subsection{Choosing consumer preferences}

We will now choose consumer preferences $c_{i\alpha}$ as follows. We assume that each specialist family has a preference for one resource class $A$ (where $A=1 \ldots  F$) with $0 \le F \le T$, and we denote the consumer coefficients for this family by $c_{i \alpha}^A$. We will also consider generalists that have no preferences, with consumer coefficients $c_{i \alpha}^{\mathrm{gen}}$. We will consider three kinds of models: one where the coefficients are drawn from Gaussian distributions, another where they are drawn from Gamma distributions (which ensure positivity of the coefficients), and finally a discrete, binary preference model.

\subsubsection{Gaussian consumer preferences}

The Gaussian model allows a continuous gradation of transporter expression levels. We assume that the variance is fixed to so that for all coefficients for all families
\begin{align}
\<(\delta c_{i \alpha}^A)^2\> = \<(\delta c_{i \alpha}^\mathrm{gen})^2\>= \frac{\sigma_c^2}{M}.
\end{align}
In the generalist family, the mean is also the same for all resources, and is given by
\begin{align}
\< c_{i \alpha}^{\mathrm{gen}} \>={\mu_c \over M}.
\end{align}
The specialist families sample from a distribution with a larger mean for resources in their preferred class:
\begin{align}
\< c_{i \alpha}^A \>=
\begin{cases}
{\mu_c \over M}\left[1+\frac{M-M_A}{M_A} q_A\right], & \text{if}\ \alpha \in A \\
{\mu_c \over M}(1-q_A), & \text{otherwise},
\end{cases}
\end{align}
where $M_A$ is the number of resources in class $A$, and $q_A$ controls how much more species from family $A$ prefer resources from class $A$. 

We have put a factor of $M$ in the denominators of the expressions for mean and variance, because the sums over $c_{i\alpha}$ in the dynamical equations (\ref{eq:dynamics}) always give rise to terms with means $M\<c_{i\alpha}\>$ or $S\<c_{i\alpha}\>$ and variances $M\<(\delta c_{i \alpha})^2\> $ or $S\<(\delta c_{i \alpha})^2\> $. The factor of $M$ allows us to keep $\sigma_c, \mu_c$ fixed when exploring the $M,S\to\infty$ limit in Appendix \ref{app:phase} below.

\subsubsection{Gamma consumer preferences}

We will also consider the case where consumer preferences are drawn from Gamma distributions, which guarantee that all coefficients will be positive. Since the Gamma distribution only has two parameters, it is fully determined once the mean and variance are specified. We parameterize the mean and variance for this model in the same way as for the Gaussian model. 

%
%
%

\subsubsection{Binary consumer preferences}

In the binary model, there are only two possible expression levels for each transporter: a low level $\frac{c_0}{M}$ and a high level $\frac{c_0}{M} + c_1$. The elements of $c_{i \alpha}^A$ are given by
\begin{align}
c_{i \alpha}^A = \frac{c_0}{M} + c_1 X_{i \alpha},
\end{align}
where $X_{i \alpha}$ is a binary random variable that equals 1 with probability
\begin{align}
p_{i \alpha}^A =
\begin{cases}
{\mu_c \over M c_1}\left[1+\frac{M-M_A}{M_A}q_A\right], & \text{if}\ \alpha \in A \\
{\mu_c \over M c_1}(1-q_A), & \text{otherwise}
\end{cases}
\end{align}
for the specialist families, and
\begin{align}
p_{i \alpha}^\mathrm{gen}=\frac{\mu_c}{Mc_1}
\end{align}
for the generalists. 

Note that the mean of the distribution is $\< c_{i\alpha}\> = p_{i\alpha} c_1 + \frac{c_0}{M}$, and the variance is $\<(\delta c_{i \alpha})^2\> =p_{i\alpha} (1-p_{i\alpha})$. Both of these scale as $1/M$ when $M\to \infty$, just like the Gaussian and Gamma versions, as long as $c_1, c_0$ and $\mu_c$ are held fixed. 

\subsection{Constructing the metabolic matrix}

We choose the metabolic matrix $D_{\alpha\beta}$ according to a three-tiered secretion model. The
first tier contains a preferred class of byproducts, such as carboyxlic acids for fermentative and respiro-fermentative bacteria, which includes $M_c$ members. The second contains byproducts of the same class as the input resource (when the import resource is not in the preferred byproduct class). For example, this could be attributed to the partial oxidation
of sugars into sugar alcohols, or the antiporter behavior of various amino acid
transporters. The third tier includes everything else. We encode this structure in $D_{\alpha\beta}$ by 
sampling each column of the matrix from a Dirichlet distribution with concentration parameters
$d_{\alpha\beta}$ that depend on the byproduct tier, so that on average a fraction $f_c$ of the secreted flux goes to the first tier, while a fraction $f_s$ goes to the second tier, and the rest goes to the third:
\begin{align}
d_{\alpha\beta} =
\begin{cases}
d_0\frac{f_c+f_s}{M_c}, & \text{if}\, \alpha = c\\
d_0\frac{1-f_c-f_s}{M-M_c}, & \text{if}\, \alpha \neq c \text{ and } \beta = c\\
d_0\frac{f_s}{M_{A(\beta)}}, & \text{if}\, \alpha,\beta \neq c \text{ and } A(\alpha) = A(\beta)\\
d_0\frac{1-f_s-f_c}{M-M_{A(\beta)-M_c}}, & \text{if}\, \alpha,\beta \neq c \text{ and } A(\alpha) \neq A(\beta).
\end{cases}
\end{align}
The parameter $d_0$ controls the randomness of the partitioning, ranging from the maximum value where all the weight is put on a single resource for $d_0 = 1$, to deterministic partitioning as $d_0 \to \infty$. 

The mean of the Dirichlet distribution is always equal to $1/M$, and the variance under this parameterization also scales as $1/M$ when the $f$'s and $d_0$ are held fixed. The sampling of $D_{\alpha\beta}$ thus following the same scaling behavior as our scheme for the consumer matrices in the $M,S\to\infty$ limit of Appendix \ref{app:phase}.

\section{Simulation and data analysis}
\label{app:procedure}
\subsection{The Community Simulator}
We implemented the above modeling framework in a Python package called ``Community Simulator,'' which can be downloaded and installed from \url{https://github.com/Emergent-Behaviors-in-Biology/community-simulator}. Once this package is installed, the data can be downloaded from \url{https://github.com/Emergent-Behaviors-in-Biology/crossfeeding-transition}, and the accompanying Jupyter notebook can be used to regenerate all the figures. The one exception is the energy flux network figure, which was generated in MATLAB using a file exported from the notebook. The repository also contains a sample MATLAB script for loading and visualizing the network file.

Community Simulator is designed to run dynamics on multiple communities in parallel, inspired by the parallel experiments commonly performed with 96-well plates. The central object of the package is a \verb+Community+ class, whose instances are initialized by specifying the initial population sizes and resource concentrations for each parallel ``well,'' along with the functions and parameters that define the population dynamics. This class contains two core methods. \verb+Propagate(T)+ sends each community to a separate CPU (for however many CPU's are available), runs the given population dynamics for a time $T$ using the SciPy function \verb+odeint+, and updates the population sizes and resource concentrations in each well to the time-evolved values. \verb+Passage(f)+ initializes a fresh set of wells by adding a fraction $f_{\mu\nu}$ of the contents of each old well $\nu$ to each new well $\mu$. (Fresh media can also be added at this point, but this feature was not relevant for the current work). The resulting values of $N_i$ are converted from arbitrary concentration units to actual population sizes using a specified scale factor, and then integer population sizes are obtained by multinomial sampling based on these values. 

The Community Simulator package also contains a set of scripts for generating models and randomly sampling parameters. \verb+MakeConsumerDynamics+ and \verb+MakeResourceDynamics+ from the \verb+usertools+ module take a dictionary of assumptions concerning the response type, metabolic regulation and resource replenishment (as described above), and generate the corresponding functions for $dN_i/dt$ and $dR_\alpha/dt$. The function \verb+MakeMatrices+, from the same module, samples the consumer matrix $c_{i\alpha}$ and the metabolic matrix $D_{\alpha\beta}$ according to the scheme described in the previous section. 

\subsection{Simulation Details}
\label{simulation}

For this paper, we generated a binary consumer matrix with $c_0 = 1$, $c_1 = 1$ and $\mu_c = 10$, and a metabolic matrix with $d_0 = 0.2$. This matrix defined a regional pool of $S = 200$ species, consuming $M = 100$ possible resource types. We used only one family and one resource class in constructing the $c_{i\alpha}$ and $D_{\alpha\beta}$ matrices (but arbitrarily assigned each resource and each species to one of four categories, as a null model for comparison with future structured simulations). We set $w_\alpha = g_i = 1$ for all $i$ and $\alpha$, and set the $l_\alpha$ for all resources equal to each other. For the multinomial sampling described above, we chose the scale factor so that $N_i = 1$ corresponds to a population of $10^6$ cells. We generated dynamics with Type-I response, no regulation, and a ``renewable'' resource replenishment model. Only resource type 0 was supplied externally, with flux $\kappa_0$, and all the other $\kappa_\alpha$'s were set to zero. 

To simulate stochastic colonization, we initialized each of 10 wells with 100 randomly chosen species from the regional pool, with a population size of $10^6$ cells per species per well. We propagated each well under Equations (\ref{eq:dynamics}) for a time $\Delta t = 11,500$, which is much longer than the maximum time required to relax to the steady state for any of the parameter regimes sampled. We used the \verb+Passage+ method with $f_{\mu\nu} = \delta_{\mu\nu}$ to periodically eliminate species whose populations became too small. For the large steady-state population sizes we consider here ($\sim 10^4 - 10^9$, see Figure \ref{fig:rankabundance} below), the multinomial sampling eliminates species whose populations are heading for extinction while minimally perturbing the dynamics of the survivors. We passaged after every 5 time units of propagation from the beginning of the simulation up to time $t = 500$, then every 100 time units until time $t = 1,500$, and finally every 1,000 time units up to the final time $t = 11,500$.

The timeseries shown in Figure 1E was generated under these assumptions, with $w_0 \kappa_0 = 500$.

We propagated these 10 initial states using this procedure for 100 different combinations of externally supplied energy flux $w_0\kappa_0$ and leakage fraction $l$, with 10 $w_0\kappa_0$ values evenly spaced on a logarithmic scale from 10 to 100, and 10 $l$ values evenly spaced from 0 to 0.9. Figure 2 of the main text shows the mean richness over the 10 parallel wells for each combination of $w_0\kappa_0$ and $l$. The richness is defined as the number of species with non-zero abundance at the end of the simulation. 

We focused on three representative examples for further analysis:
\begin{enumerate}
	\item {\bf Syntrophy-Limited}: $w_0\kappa_0 = 1000$, $\langle l_\alpha \rangle = 0.1$
	\item {\bf Energy-Limited}: $w_0\kappa_0 = 28$, $\langle l_\alpha \rangle = 0.6$
	\item {\bf Similarity-Limited}: $w_0\kappa_0 = 1000$, $\langle l_\alpha \rangle = 0.9$.
\end{enumerate}
The rank-abundance plots in Figure 2 of the main text show the population sizes in all 10 wells from each of these examples, after normalizing them by the total biomass $\sum_i N_i$. The plots were truncated at a relative abundance of 0.5\% for clarity. Rank-abundance plots for these same three examples in absolute units with no truncation can be found in Figure \ref{fig:rankabundance} below.

\subsection{Susceptibilities}
One important property of an ecosystem is its sensitivity to changes in environmental conditions. Figure 3 of the main text quantifies this sensitivity in terms of a set of susceptibilities, defined by
\begin{align}
\chi_{\alpha\beta} &\equiv \frac{\partial \bar{R}_\alpha}{\partial \kappa_\beta}\\
\eta_{i\beta} &\equiv \frac{\partial \bar{N}_i}{\partial \kappa_\beta}
\end{align}
where $\bar{N}_i$, $\bar{R}_\alpha$ are the steady-state consumer populations and resource concentrations, respectively. 

For the case of externally supplied resources and Type-I growth, setting Equations (\ref{eq:dynamics}) equal to zero and differentiating with respect to $\kappa_\beta$ yields:
\bea
0&=& \sum_\alpha  (1-l_\alpha) w_\alpha c_{i \alpha} \chi_{\alpha\beta} \\
-\tau_\alpha^{-1} \delta_{\alpha\beta} &=& \sum_{\gamma} \left( \sum_j c_{j \gamma} N_j \left[ D_{\alpha \gamma}{w_\gamma \over w_\alpha} l_\gamma - \delta_{\gamma\alpha} \right]  - \tau_\alpha^{-1} \delta_{\gamma\alpha}\right)  \nonumber \\
&&\times \chi_{\gamma\beta}+  \sum_{j \gamma} c_{j \gamma}  \left[ D_{\alpha \gamma}{w_\gamma \over w_\alpha} l_\gamma - \delta_{\gamma\alpha} \right] R_\gamma \eta_{j\beta}
\eea
The last equation can be reorganized as 
\begin{align}
-\tau_\alpha^{-1} \delta_{\alpha\beta} &= \sum_{\gamma} \bigg(\sum_j c_{j \gamma} N_j \left[ D_{\alpha \gamma}{w_\gamma \over w_\alpha} l_\gamma - \delta_{\gamma\alpha} \right]  \nonumber\\
&- \tau_\alpha^{-1} \delta_{\gamma\alpha}\bigg) \chi_{\gamma\beta} \nonumber \\
&+  \sum_{j \gamma} c_{j \gamma}  \left[ D_{\alpha \gamma}{w_\gamma \over w_\alpha} l_\gamma - \delta_{\gamma\alpha} \right] R_\gamma \eta_{j\beta}
\end{align}
For each value of $\beta$, this system of linear equations can be solved for $\chi_{\gamma\beta}$ and $\eta_{j\beta}$ by simply inverting a matrix (once the terms corresponding to extinct species have been removed). 

The histograms of Figure 3D in the main text contain the diagonal elements $\chi_{\alpha\alpha}$ for all resources except for the one supplied externally ($\alpha = 0$), which might be expected to behave somewhat differently. The $\chi_{\alpha\alpha}$ values from all 10 parallel communities are included in the histogram. We generated one histogram for the similarity-limited regime, and one for the energy-limited regime, using the examples defined in Section \ref{simulation} above.

\subsection{Niche Overlap}
To find out what controls the diversity of the diverse regime, we varied the niche overlap, which quantifies the similarity among consumer preferences within the regional species pool. We did this by holding $\mu_c$ fixed, and varying $c_1$ from its original value of 1 down to a minimum value of 0.12. For each value of $c_1$, we generated 10 $c_{i\alpha}$ matrices, which each defined a regional pool of 200 species. We then repeated the procedure of Section \ref{simulation} above for each of these regional pools: initializing 10 wells with 100 species and running them to the steady state with the same sequence of propagation and passage steps. The final richness of each community is plotted in Figure 4 of the main text as a translucent point, such that more common richness values are darker. We included all three examples defined at the end of Section \ref{simulation} in the plot, and colored both examples from the resource-limited regime blue, while the diverse regime was colored red.

\subsection{Beta Diversity}
To examine the beta diversity patterns in each regime, we initialized 200 wells with 100 randomly chosen species from the regional pool of 200 species, and propagated them to steady state following Section \ref{simulation} under the three different choices of $w_0\kappa_0$ and $l$ listed at the end of that section.  To visualize the variation among these communities, we used the Python package scikit-learn \cite{scikit-learn} 
to compute the first two principle components of the set of composition vectors in each regime. We then projected the compositions onto the plane spanned by these vectors, and generated a scatter plot of the results. We also computed the percentage of the total variance accounted for by each of these two principal components, and indicated the value in parentheses on each axis.

\subsection{Data Format}
The output of all the simulations was saved to a set of Microsoft Excel spreadsheets, which can be easily imported into Python for analysis using the Pandas package. Each simulation generated four files: final consumer populations (`\verb+Consumers+'), final resource concentrations (`\verb+Resources+'), a metadata summary (`\verb+Parameters+'), and initial conditions (`\verb|Initial_State|'). The $c_{i\alpha}$ and $D_{\alpha\beta}$ matrices as well as the $m_i$ and $w_\alpha$ were pickled into a binary file (`\verb+Realization+'). The file names also include the date on which the data was generated, and a task ID when multiple files were generated on the same day. 

The first column in the consumer and resource tables is the index of the simulation run. The second and third columns of the consumer file are the family ID and species ID, respectively. In the resource file, these columns contain the class ID and resource ID. The remaining columns contain the populations/concentrations for each well. The consumer populations are in units of $10^6$ cells. 

All the parameters that change between runs are included in the metadata file (`\verb+Parameters+'). The first column of this file is the simulation run index, corresponding to the index in the consumer and resource files. 

The initial conditions file contains the initial population sizes for each of the wells.


\section{Robustness of qualitative results}
\label{app:robust}
In this Appendix, we test the robustness of our qualitative results by modifying the modeling assumptions in five ways. We have given each way a descriptive name, which can be used to look up the raw data files from the supplemental data folder using the \verb+file_list.csv+ table:
\begin{itemize}
	\item \verb+main_dataset+ is the data from the main text
	\item \verb+type_II+ uses a Type II functional response, with $K = 20$.
	\item \verb+dense_metabolism+ has a dense metabolic matrix with $d_0 = 0.001$.
	\item \verb+randomness+ adds (quenched) random variation to $w_\alpha$ and $l_\alpha$, with standard deviations 0.1 and 0.03, respectively.
	\item \verb+Gaussian_sampling+ samples the $c_{i\alpha}$'s from Gaussian distributions, with the same mean $0.11$ and standard deviation $0.3$ as the binary matrix used in the main text.
	\item \verb+Gamma_sampling+ samples the $c_{i\alpha}$'s from Gamma distributions, with the same mean and variance.
\end{itemize}

\begin{figure*}
	\includegraphics[width=18cm,trim=50 -20 0 0]{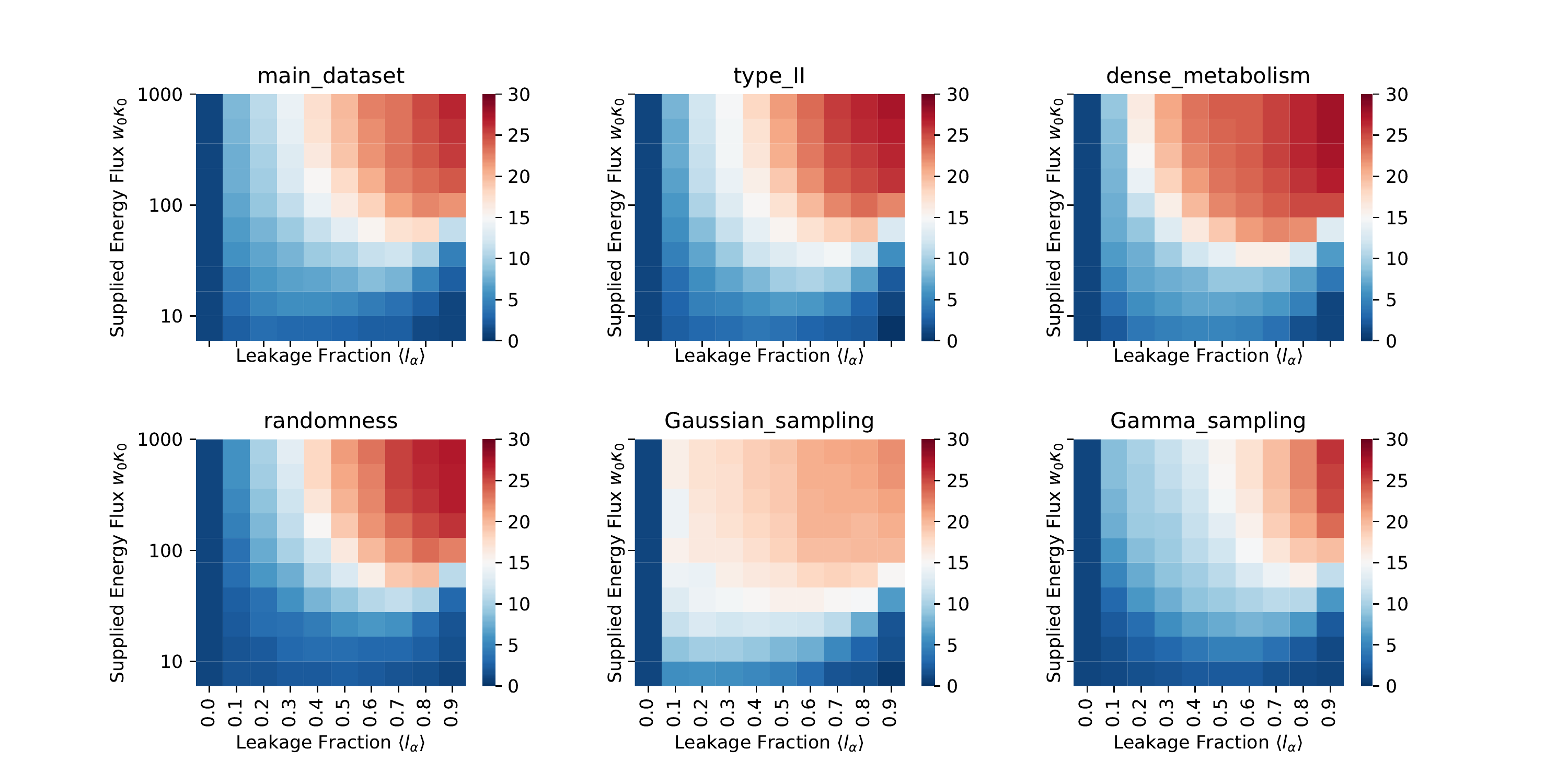}
	\caption{{\bf Richness vs. $w_0\kappa_0$ and $\langle l_\alpha\rangle$} We generated 200 species, initialized 10 communities of 100 species each from this pool, and ran the dynamics to steady state under different combinations of  $w_0\kappa_0$ and $\langle l_\alpha\rangle$, for each of the six model choices listed at the end of Section \ref{simulation}. The color of each square indicates the mean number of non-extinct species at the end of the simulation, over all 10 communities at each combination of $w_0\kappa_0$ and $\langle l_\alpha\rangle$.}
	\label{fig:heatmap2}
\end{figure*}

\begin{figure*}
	\includegraphics[width=18cm,trim=50 -20 0 0]{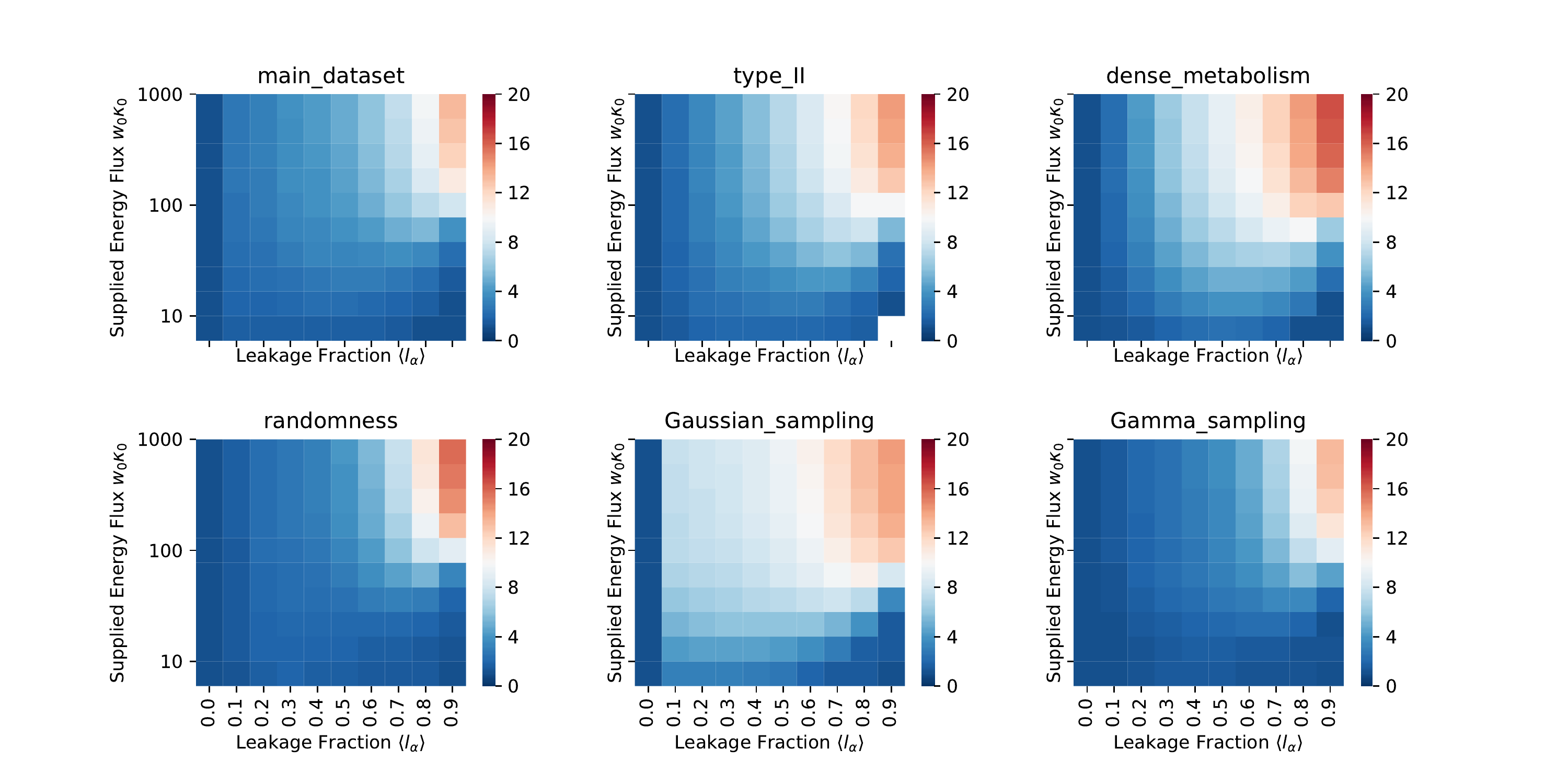}
	\caption{{\bf Simpson Diversity vs. $w_0\kappa_0$ and $\langle l_\alpha\rangle$}. Simpson Diversity was computed according to Equation (\ref{eq:simpson2}), using the same data as Figure \ref{fig:heatmap2}.}
	\label{fig:simpson}
\end{figure*}

\begin{figure*}
	\includegraphics[width=16cm,trim={1cm 1.5cm 0 0},clip]{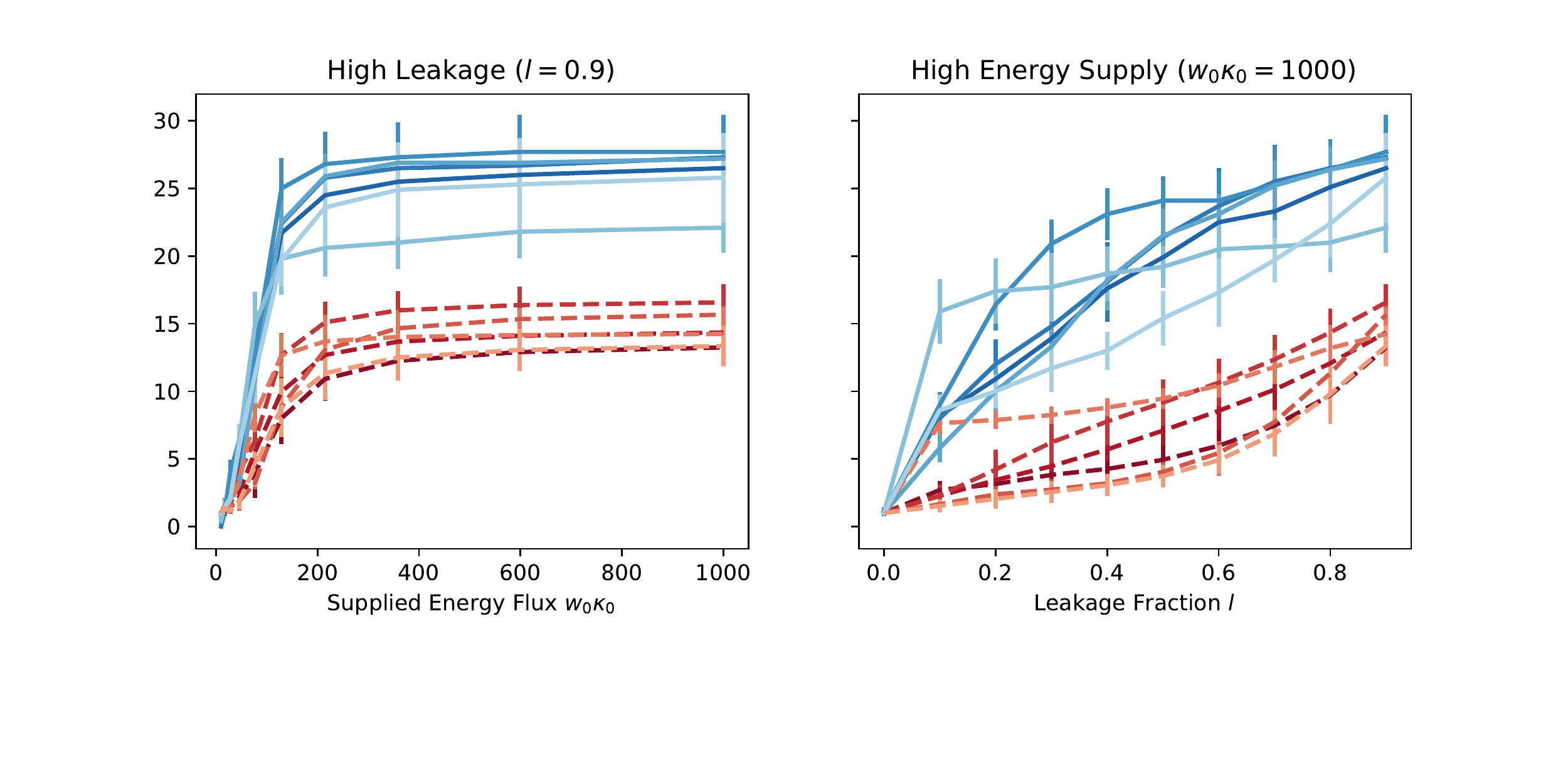}
	\caption{{\bf Richness (blue solid) and Simpson Diversity (red dotted) for cuts through the heat map.} All six modeling choices are plotted, going from dark to light in the order listed in the text. Error bars are plus and minus one standard deviation, where the standard deviation is computed over the 10 parallel communities at each set of parameter values.}
	\label{fig:simpsonrichness}
\end{figure*}

The following sections describe each of these choices in more detail. Figures \ref{fig:heatmap2}, \ref{fig:rankabundance}, \ref{fig:fluxdiversity} and \ref{fig:susceptibility} show the key plots from the main text along with the new versions generated under all these modified assumptions. Figures \ref{fig:simpson} and \ref{fig:simpsonrichness} display another diversity measure not discussed in the main text: the Simpson Diversity (S. D.). This is defined analogously to the ``effective number of resources consumed'' presented in Equation (6) of the main text:
\begin{align}
\label{eq:simpson2}
{\rm S. D.} = \left[ \sum_i \left(\frac{N_i}{N}\right)^2\right]^{-1}
\end{align}
where $N \equiv \sum_i N_i$. As discussed in the main text in connection with resource fluxes, this quantity approaches 1 when there is one large $N_i \approx N$ and all the other populations are very small. It approaches the number of species (i.e., the richness) as the biomass distribution becomes more uniform.

\subsection{Type-II Growth}
We chose the Monod parameter $K = 20$ in the Type-II growth simulations in order to ensure that at least one species would survive in the steady state in all simulations. The maximum possible incoming energy flux in the Type-II model is equal to $0.1 K$ when $w_\alpha = 1$ and $l = 0.9$, and this must exceed $m_i \approx 1$ for a species to survive. $K = 20$ provides a maximum flux of 2 in this case.

\subsection{Metabolic Matrices}
The metabolic matrices $D_{\beta\alpha}$ are plotted in Figure \ref{fig:network} of the main text for \verb+main_dataset+ and \verb+dense_metabolism+ (all other simulations use the same metabolic parameters as \verb+main_dataset+). We see that $d_0 = 0.2$ leads to a very sparse matrix, with only a few secreted byproducts per input resource, while the secretion fractions for $d_0 = 0.001$ are much more uniform.

\subsection{Randomness in $w_\alpha$ and $l_\alpha$}
To relax the assumption of all the $w_\alpha$'s and $l_\alpha$'s being equal, we sampled these two vectors from Gaussian distributions. We chose the standard deviations of the distributions to be small enough that both quantities would almost always be positive, and $l_\alpha$ would remain less than 1.

\subsection{Gaussian and Gamma Sampling}
Sampling consumer preferences from the continuous Gaussian and Gamma distributions makes the differential equations much more stiff than in the binary case. To ensure stable operation of the integrator, we ``passaged'' the cells every 0.1 time units. Each call of the ``passage'' method zeros out small negative values of resource concentration or consumer population that arise because of numerical error, in addition to setting small consumer populations to zero. This high frequency of passaging made the simulation more computationally intensive, so we only propagated these simulations for 200 time units. We computed the root-mean-square difference between the per-capita growth rates $(1/N_i)(dN_i/dt)$ and zero to check whether the simulations had converged. We found that all of them had acceptably converged, except for some of the runs at $w_0\kappa_0 < 100$ in the Gaussian case. The Gaussian model is unphysical, because almost half of the consumer preferences are less than 0 for these simulations, and so we decided not to spend more computation time in pursuit of convergence.

\subsection{Quantification of Nestedness}
\label{sec:nested}
Almeida-Neto \emph{et al.} have introduced a quantitative measure of nestedness, called the ``Nestedness metric based on Overlap and Decreasing Fill,'' or NODF \cite{almeida2008consistent}. Figure \ref{fig:nestedness-test} shows how the NODF depends on the relative abundance threshold for the Tara Oceans data, as compared with two null models taken from the Earth Microbiome Project analysis \cite{EMP}. Null Model 1 keeps the richness of each sample the same while randomly altering the identities of the surviving species. It tells us how much nestedness we should expect by chance from a set of samples with the given levels of diversity, in the absence of any ecological mechanisms. Null Model 2 keeps the prevalence of each species the same while randomly assigning it to different samples. This procedure generates another well-defined family of random matrices with similar bulk statistics to the original data, but lacks the operational interpretation of Null Model 1. For this reason, the comparison with Null Model 1 is more meaningful, but we include Null Model 2 for completeness. 

For each of the null models and each value of the relative abundance threshold, we generated 100 random permutations of the data matrix and computed the mean and standard deviation of their NODF scores. We found that the actual nestedness exceeds that of Null Model 1 by at least 20 standard deviations for all values of the threshold. For the relative abundance threshold of 0.5\% employed in Figure \ref{fig:data}, the true NODF also exceeds Null Model 2 by 5.8 standard deviations.

The figure also shows histograms generated with the two null models from the simulation data of Figure \ref{fig:beta}A. The actual NODF (=0.46) is more than 100 standard deviations above the mean nestedness for both models.

To compute the NODF, we employed the following algorithm, which is implemented in the Community Simulator package (in the \verb+analysis+ module). Let $n$ be the number of columns, and $m$ the number of rows in a matrix $A$. Let $D_c$ be an $n\times n$ matrix, such that $(D_c)_{ij} = 1$ if the sum of column $i$ is greater than the sum of column $j$, and 0 otherwise. Similarly, let $D_r$ be an $m\times m$ matrix such that $(D_r)_{ij} = 1$ if the sum of row $j$ is greater than the sum of row $i$, and zero otherwise. Let $B$ be the row-normalized matrix, where each row of $A$ has been divided by the sum over the row. And let $C$ be the column-normalized matrix, where each column has been normalized by the sum over the column. Then the NODF of the matrix $A$ is
\begin{align}
{\rm NODF} = 2\frac{{\rm Tr}(A^T D_r B) + {\rm Tr}(A D_c C^T)}{n(n-1)+m(m-1)}
\end{align}
where ${\rm Tr}$ represents the trace operation.

\section{Phase Transition}
\label{app:phase}

A phase transition in physics is characterized by a discontinuous change in the value of an observable or its derivative as an intensive parameter is varied, in the ``thermodynamic limit'' of infinite system size. In an ecological context, the analog to system size is the number of possible resource types $M$, or the initial number of species $S$. Several recent works have explored the analytic computations that become tractable in the $M,S \to\infty$ limit of various models, while $\gamma\equiv M/S$ remains constant (when the model is resource-explicit) \cite{Advani2018,Tikhonov2017,Bunin2017,Barbier2018}. Taking this limit requires several decisions about how to scale the rest of the parameters. Our sampling scheme for the $c_{i\alpha}$ and $D_{\alpha\beta}$ matrices, described in Appendix \ref{app:procedure}, follows the canonical strategy for studying spin glasses, where the random coupling parameters $J_{ij}$ are chosen such that the mean and variance are both proportional to $1/M$ \cite{Nishimori}. The maintenance costs $m_i$, on the other hand, are sampled from the same distribution regardless of the value of $M$. Finally, we note that the total amount of energy $w_0 \kappa_0$ supplied to the system is an ``extensive'' parameter, and that the scaling analysis should be performed with the corresponding ``intensive'' parameter $w_0\kappa_0/M$ held fixed.

Figure \ref{fig:diversity-scaling} shows how the consumer richness scales with $M$ for each of the three examples discussed in the main text. The first two examples come from the resource-limited regime. The ``syntrophy-limited'' example has $w_0\kappa_0/M = 10, l = 0.1$, while the ``energy-limited'' example has $w_0\kappa_0/M = 0.28, l = 0.6$. The third, ``similarity-limited'' example comes from the diverse regime, with $w_0\kappa_0/M = 10, l = 0.9$. The richness appears to scale like $M^\alpha$ with exponent $\alpha < 1$ for the resource-limited examples, and $\alpha = 1$ for the diverse example. If this scaling holds asymptotically as $M\to\infty$, then the system exhibits a true phase transition, with the normalized richness (richness/$M$) vanishing in the resource-limited regime, while remaining finite in the diverse regime. The gray line in the right-hand panel illustrates what this would look like, with a discontinuity in the derivative of the normalized richness as a function of $l$ or $w_0\kappa_0/M$. This evidence is by no means conclusive, since we only have access to a single decade of $M$ values. To reach three decades of $M$ values would require solving 40,000 coupled ODE's involving matrices $c_{i\alpha}$ and $D_{\alpha\beta}$ with size $20,000\times 20,000$. Each matrix would thus have $4\times 10^8$ entries, corresponding to 50 MB of memory for binary entries or 800 MB of memory for floating-point entries. This computation is feasible but non-trivial, demanding significantly more attention to how the matrix multiplications are implemented and how the matrices are passed around in memory. We are currently working on an update to the Community Simulator package that implements the core computations in PyTorch, which enables GPU acceleration of matrix multiplication and may allow for calculations on this scale.

For completeness, Figure \ref{fig:diversity-scaling2} shows how four other natural observables scale with system size. These are the Simpson and Shannon diversity of the steady-state consumer and resource abundances \cite{Tuomisto2010}. To compute these quantities, one first obtains relative abundances $f_i = N_i/\sum_i N_i$ and $f_\alpha = R_\alpha/\sum_\alpha R_\alpha$. In terms of these fractions, the Simpson diversity is
\begin{align}
D_{\rm Sim} = \left( \sum_i f_i^2 \right)^{-1}
\end{align}
and is related to the Inverse Participation Ratio commonly analyzed in spin glass problems, while the Shannon diversity is
\begin{align}
D_{\rm Sh} = \exp\left( -\sum_i f_i \ln f_i\right),
\end{align}
and is simply the exponential of the Shannon entropy of the distribution (where the sum is taken only over the species with nonzero abundance). Both of these quantities are equal to 1 in the limit where a single type dominates the distribution, and equal the total number of surviving types when all the types have the same abundance. The Simpson and Shannon diversity of the consumers appear to saturate at a finite values in the large $M$ limit of the resource-limited regime. When measured in these ways, the diversity of this regime thus appears to be insensitive to the size of the regional species pool and to the number of possible resource types, and is controlled by the energy supply, leakage fraction, and probably also the sparsity of the $D_{\alpha\beta}$ matrix.

\begin{figure*}
	\includegraphics[width=13cm,trim={0 4cm 0 4cm},clip]{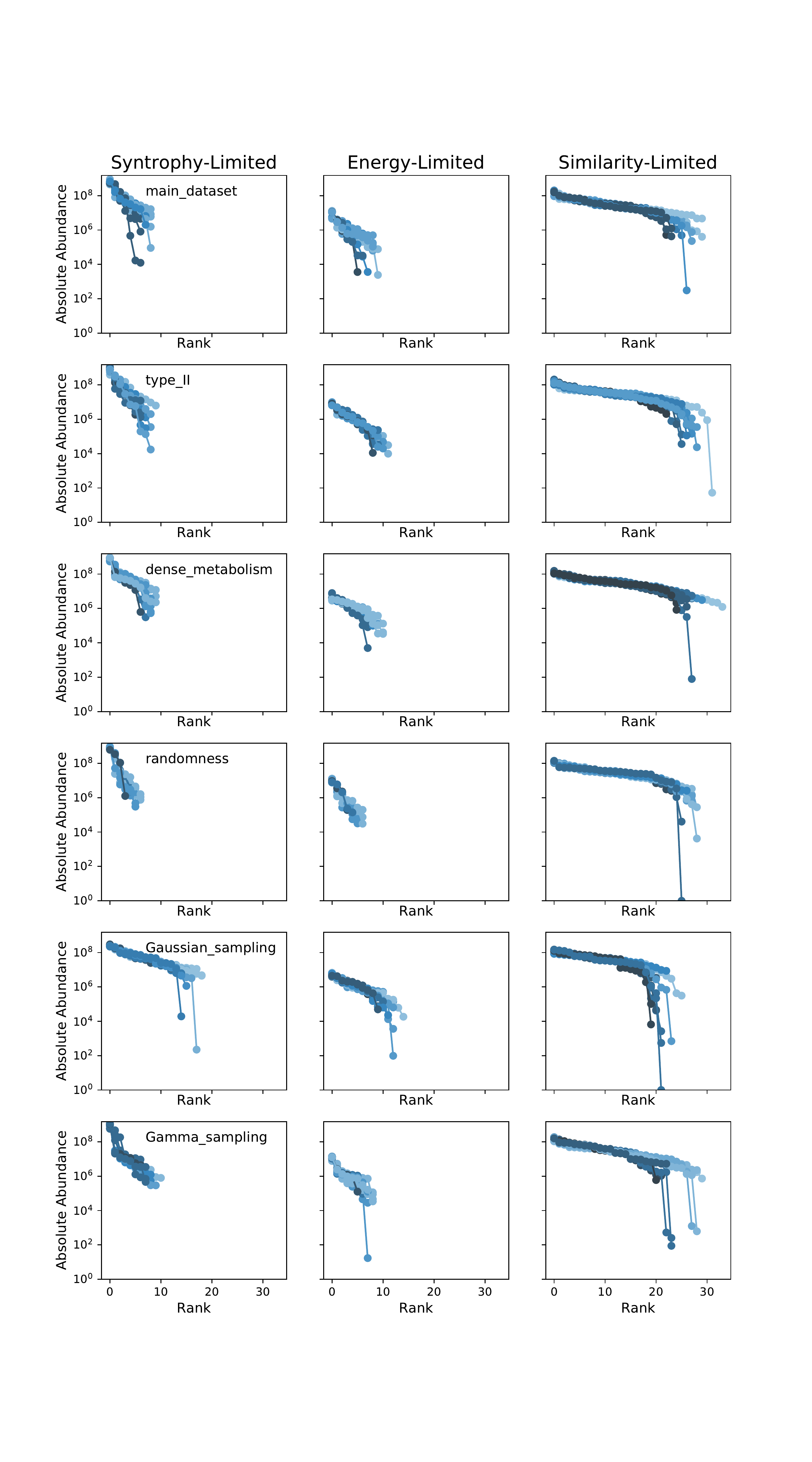}
	\caption{{\bf Rank-abundance curves for three representative examples.} The abundance is plotted in absolute number of cells, as opposed to the relative abundance of the main text, and includes all species, with no truncation. The lower limit of the vertical axis is set to 1 cell, which is the smallest possible population size once the Passage method described in Appendix \ref{app:procedure} has generated integer population values. The three examples are the same ones highlighted in Figure 2 of the main text, and listed in Section \ref{simulation} above.}
	\label{fig:rankabundance}
\end{figure*}

\begin{figure*}
	\includegraphics[width=18cm,trim={50 0 0 0},clip]{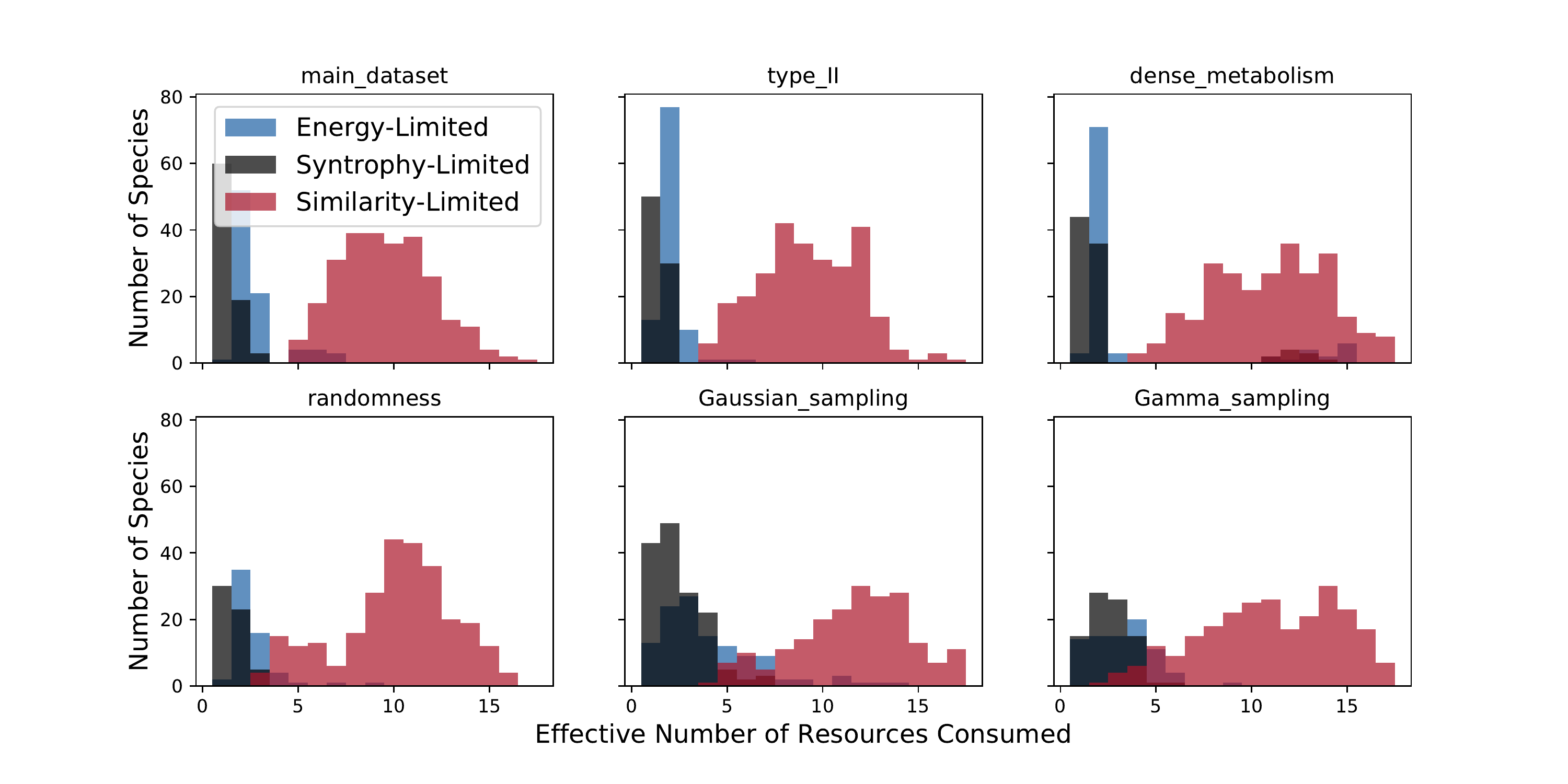}
	\caption{{\bf Effective number of resources consumed.} The effective number of resources consumed $M^{\rm eff}_i$ is computed as described in the main text for the same three examples. The $y$ axis indicates the total number of species falling into each bin from the combination of all 10 parallel communities in each example.}
	\label{fig:fluxdiversity}
\end{figure*}

\begin{figure*}
	\includegraphics[width=18cm,trim={50 0 0 0},clip]{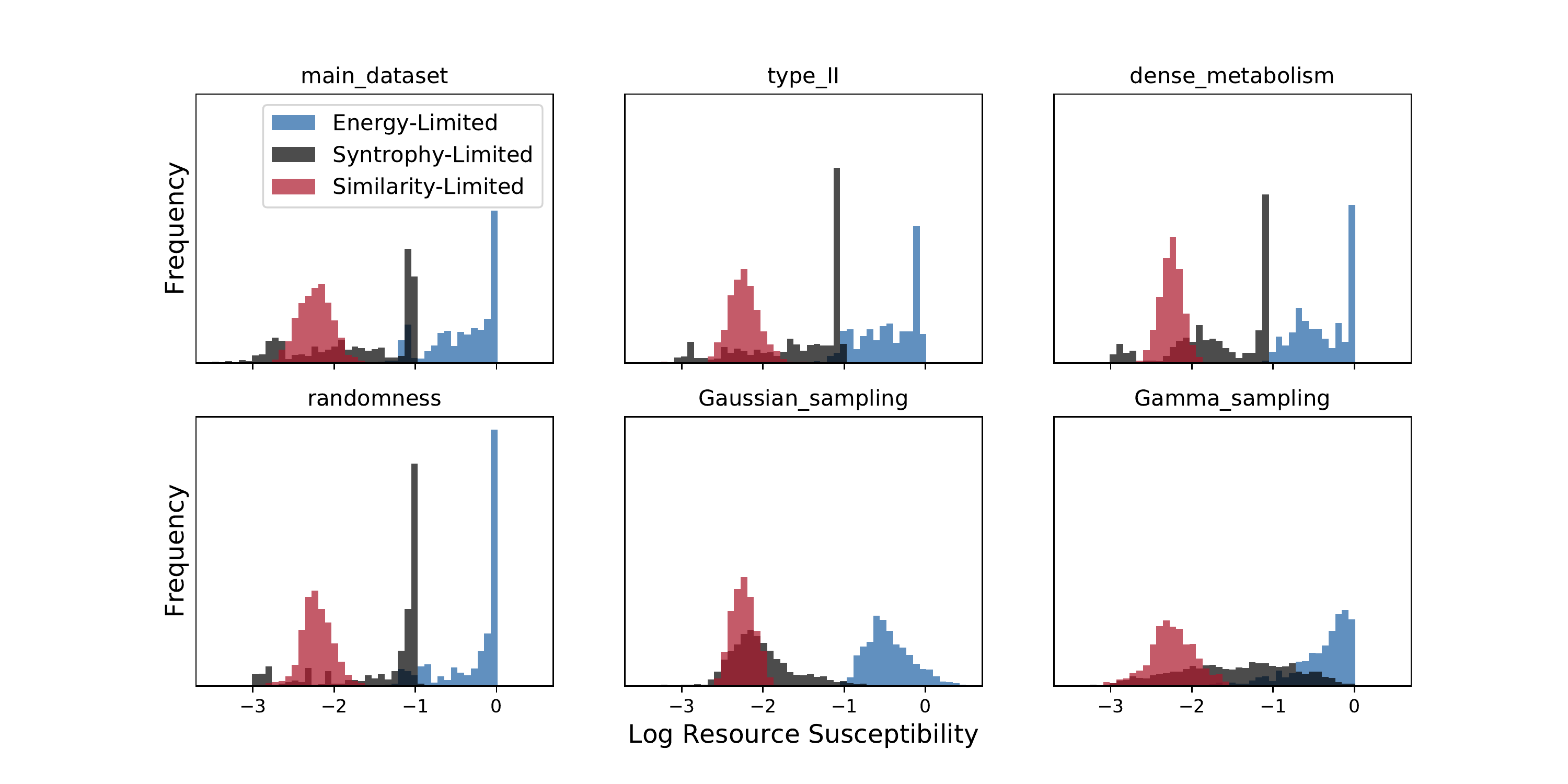}
	\caption{{\bf Resource susceptibility.} Histograms of the logarithm $\log_{10} \partial \bar{R}_\alpha/\partial \kappa_\alpha$ are plotted for the same three examples. The $y$ axis indicates the total number of species falling into each bin from the combination of all resources for all 10 parallel communities in each example, excepting the externally supplied resource $\alpha = 0$.}
	\label{fig:susceptibility}
\end{figure*}

%

\begin{figure*}
	\includegraphics[width=13cm,trim=0 -20 0 0]{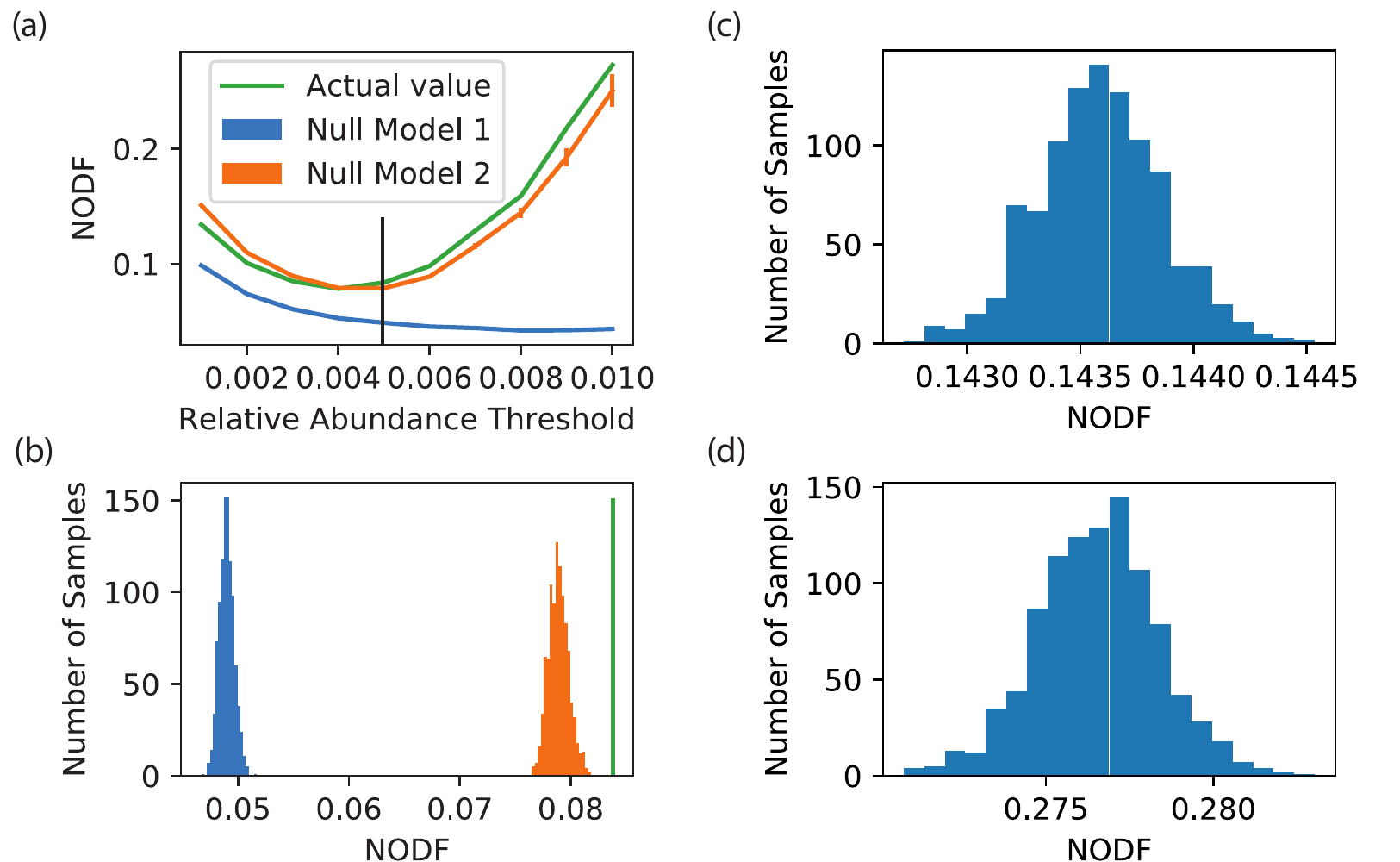}
	\caption{{\bf Quantification of nestedness}.  (a) Nestedness as quantified by NODF \cite{almeida2008consistent} of the Tara Oceans data, for different values of the relative abundance threshold. See Appendix \ref{sec:nested} for the full quantification algorithm. Also shown are the mean NODF over 100 samples from two null models, with error bars representing $\pm$1 standard deviation. Null Model 1 keeps the richness of each sample the same while randomly altering the identities of the surviving species. Null Model 2 keeps the prevalence of each species the same while randomly assigning it to different samples. (b) Histograms of 1000 samples from each null model at the relative abundance threshold of 0.005 used in the main text, and indicated with a black line in the first panel. (c), (d) Null models for comparison with the NODF value of 0.46 obtained for the simulation data. The two panels show histograms of 1,000 samples from null models 1 and 2, respectively.} 
	\label{fig:nestedness-test}
\end{figure*}

\begin{figure*}
	\includegraphics[width=16cm,trim=0 -20 0 0]{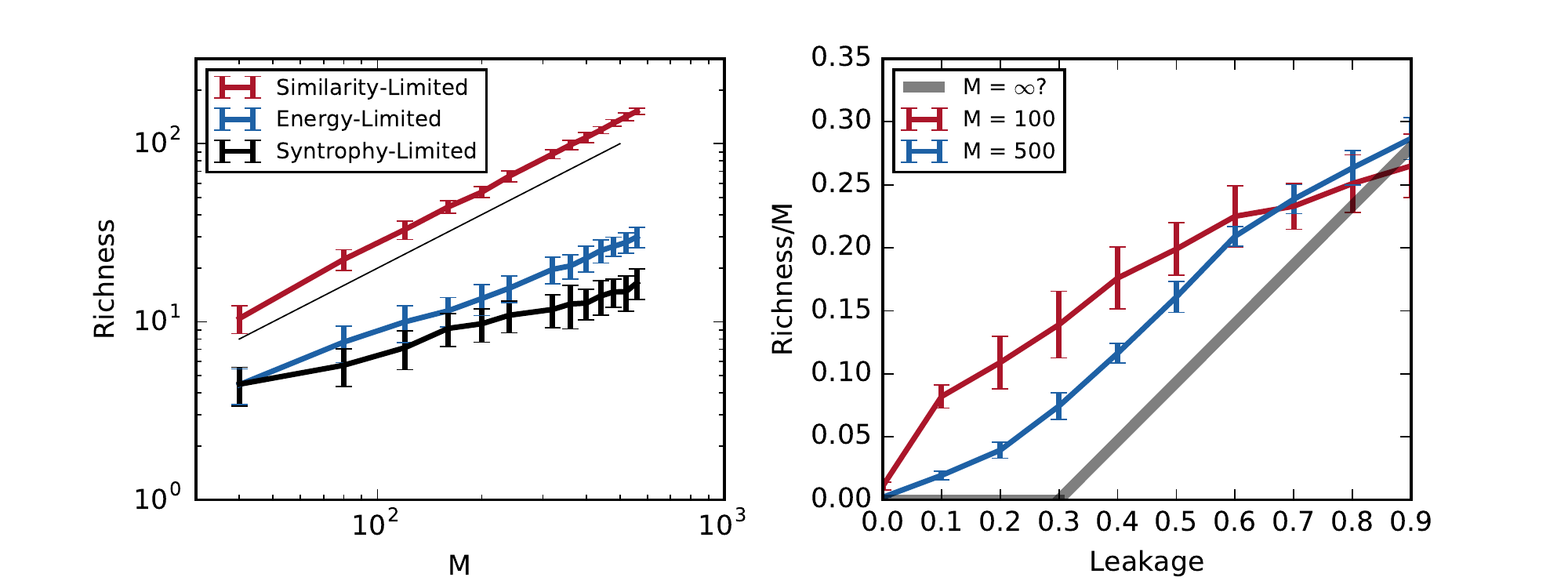}
	\caption{{\bf Scaling of consumer richness with system size}. Left: Community richness in the three examples of Section \ref{simulation} as a function of the number of resource types $M$ (with the rest of the parameters scaled as described in Appendix \ref{app:procedure}). The thin black line shows the slope corresponding to a linear scaling with $M$. Error bars are standard deviations over 100 samples per point. Right: Normalized richness as a function of leakage fraction at high energy supply $w_0\kappa_0/M = 10$. The top red line comes from the simulations of Figure 2 of the main text ($M = 100$), and the middle blue line comes from new simulations with $M = 500$. The bottom gray line illustrates a possible $M\to \infty$ limit, which would correspond to a continuous phase transition. Error bars are standard deviations over 10 samples per point.}
	\label{fig:diversity-scaling}
\end{figure*}

\begin{figure*}
	\includegraphics[width=18cm,trim=50 40 0 0]{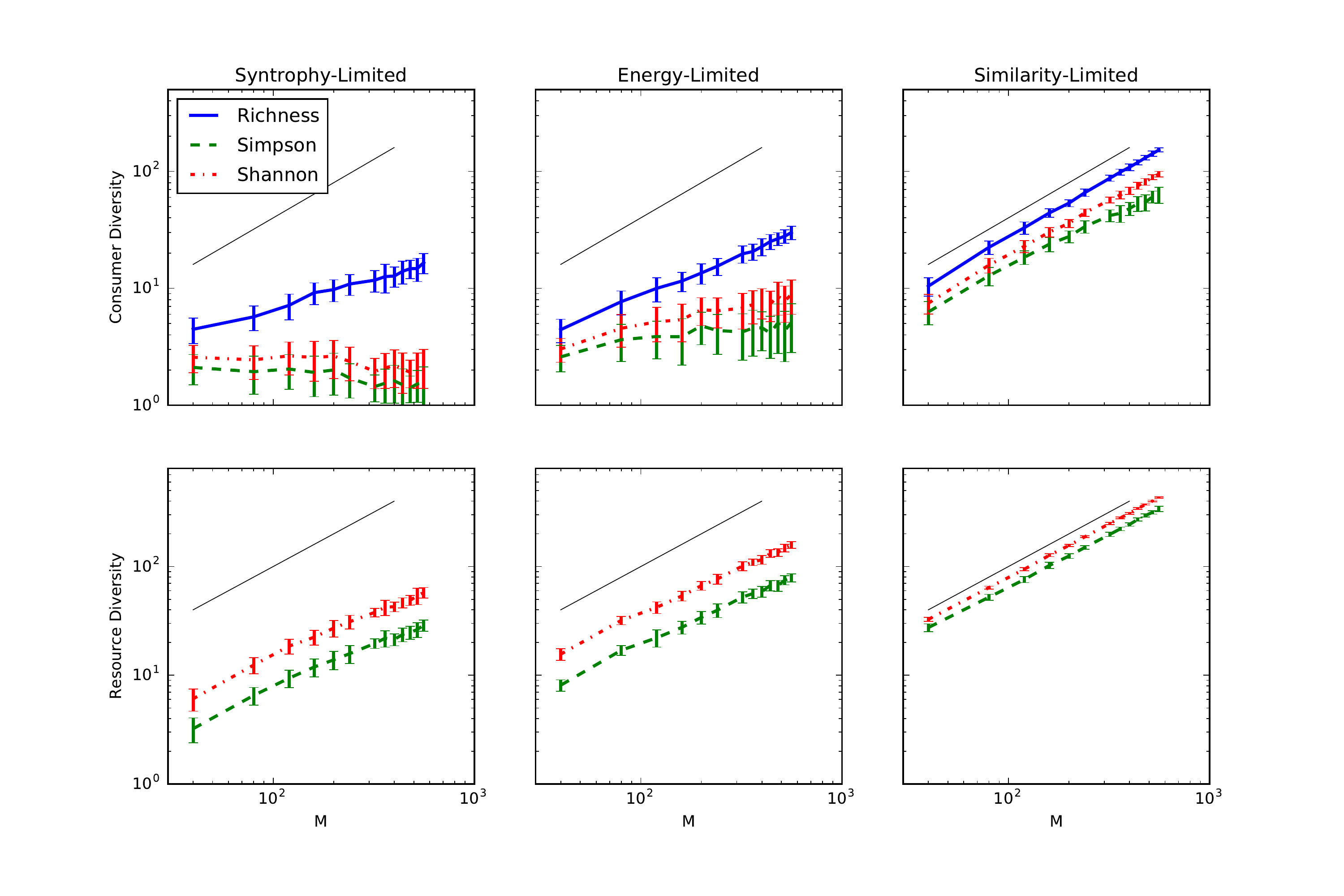}
	\caption{{\bf Scaling of other observables.} Diversity of steady-state consumer and resource abundances in the three examples of Section \ref{simulation}, as a function of the number of resource types $M$ (with the rest of the parameters scaled as described in Appendix \ref{app:procedure}). The thin black line shows the slope corresponding to a linear scaling with $M$. Error bars are standard deviations over 100 samples per point.}
	\label{fig:diversity-scaling2}
\end{figure*}

\begin{figure*}
	\includegraphics[width=16cm,trim=0 20 0 0]{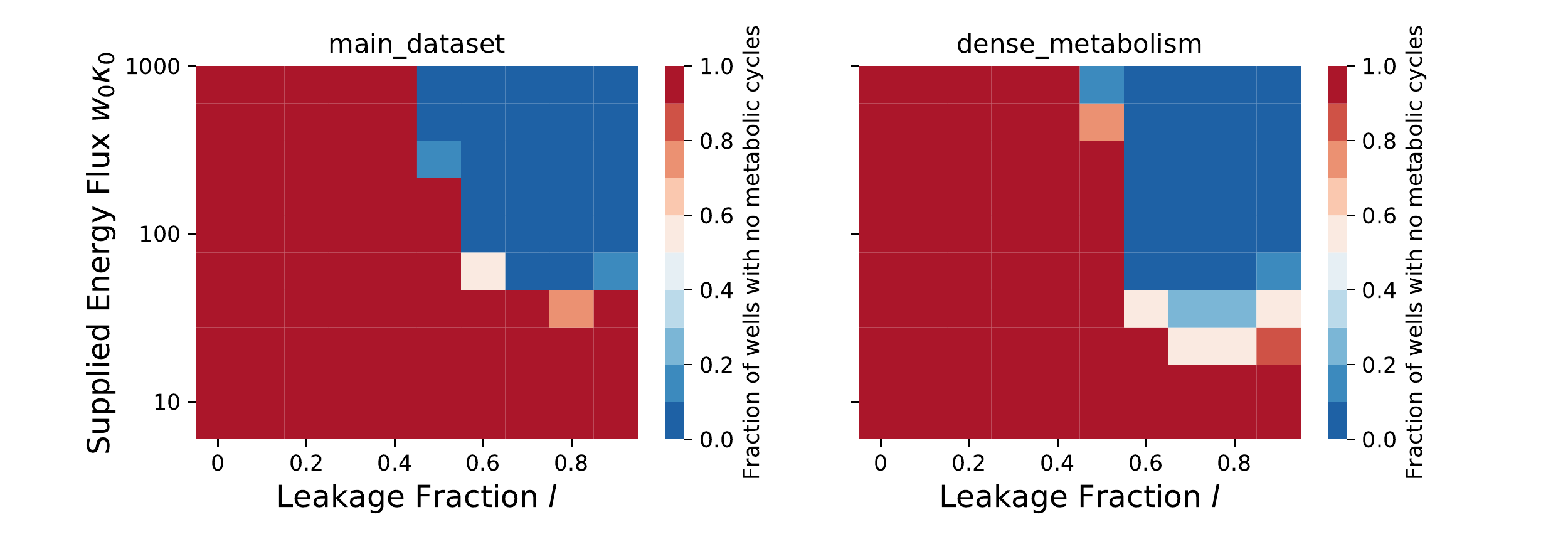}
	\caption{{\bf Flux network topology.} We generated flux networks like those of Figure \ref{fig:network} for both the sparse (`main\_dataset') and the dense (`dense\_metabolism') metabolic matrices shown there. We pruned the graphs as in the main text figure by removing edges with flux less than 1\% of maximum flux in the network. For each condition, we determined how many of the ten replicates had acyclic graphs, which could thus be topologically ordered like the left-hand panels of the figure. The color of each square represents the fraction of acyclic graphs for the given values of $l$ and $w_0\kappa_0$.}
	\label{fig:dag}
\end{figure*}

\begin{figure*}
	\includegraphics[width=10cm,trim=0 20 0 0]{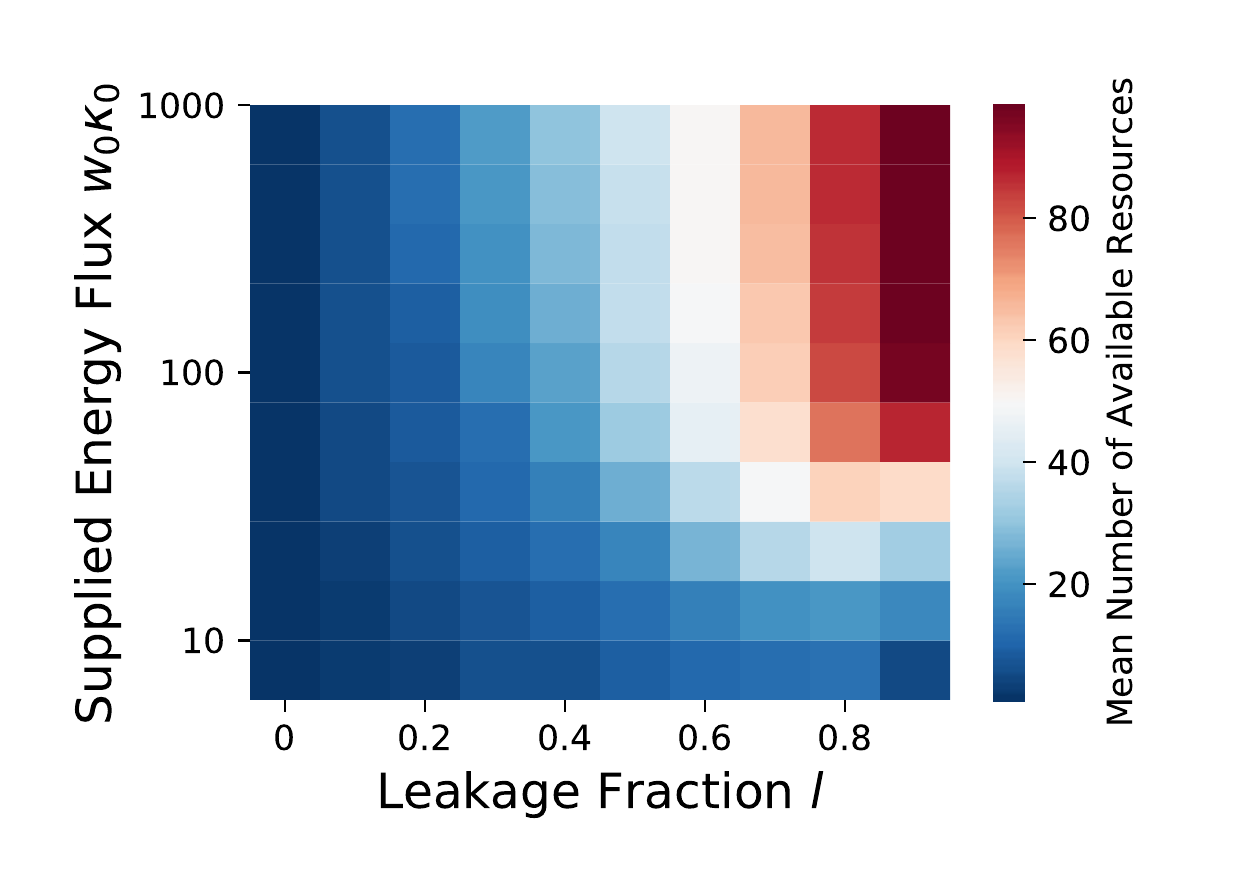}
	\caption{{\bf Number of available resources.} The mean number of resources (over all ten initial communities) with abundances $R_\alpha$ above 0.1 is plotted for each of the conditions summarized in Figure \ref{fig:heatmap}. This threshold is the level at which an average consumer equally consuming 10 resources would be able to satisfy its full maintenance cost.}
	\label{fig:resources}
\end{figure*}

\begin{figure*}
	\includegraphics[width=16cm,trim=0 30 0 0]{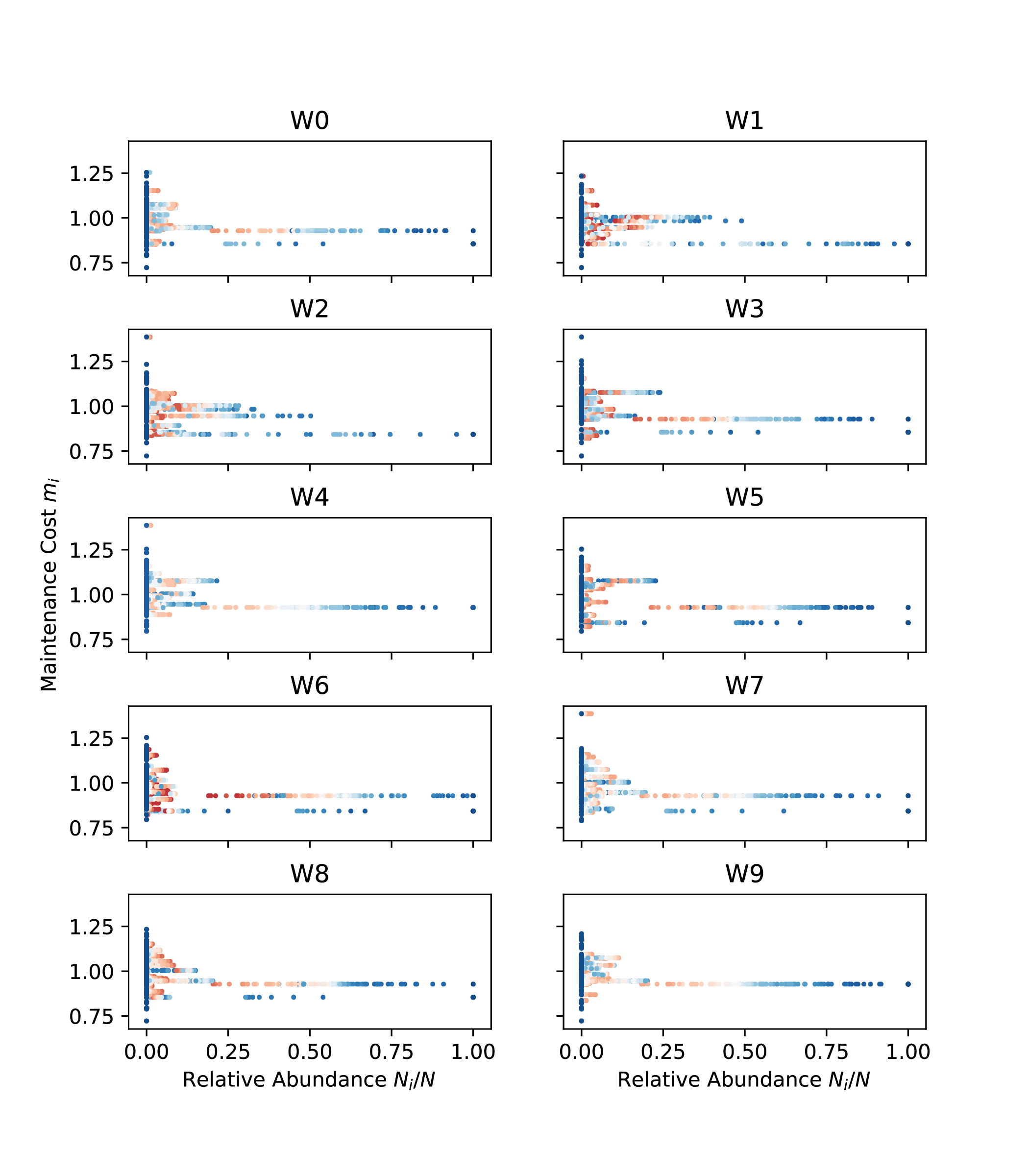}
	\caption{{\bf Maintenance costs $m_i$.} The maintenance cost $m_i$ for each species is plotted against the relative abundance of that species in its community, for each of the simulations in Figure \ref{fig:heatmap}.  The ten panels contain data for each of the ten initial pools of species. In each panel, each species has 100 data points, one for each of the 100 combinations of $l$ and $w_0\kappa_0$. The points are colored by the richness of the steady-state community, with blue being least diverse and red the most diverse. Every point in a given panel corresponds to a species that was initially present in the community, so points on the $m_i$ axis where all the relative abundances vanish are species that never survive to steady state in any of the sampled conditions.}
	\label{fig:mi}
\end{figure*}

\end{document}